\documentclass[iop,revtex4]{emulateapj}
\usepackage{natbib}
\usepackage{color}
\usepackage{bm}
\usepackage{hyperref}
\usepackage{float}
\usepackage{latexsym}
\usepackage{amsmath}
\usepackage{amssymb}
\usepackage{rotate}
\usepackage{epstopdf}
\usepackage{graphicx}
\usepackage{url}
\usepackage{booktabs} 
\usepackage[titletoc,title]{appendix}
\usepackage{longtable}
\usepackage{lscape}

\include{epsf}
\DeclareGraphicsExtensions{.eps}

\begin{document}

\def \sb {mag\,arcsec$^{-2}$}
\def \mue {$<\!\mu\!>_{e}$}
\def \ellip {$\overline{E}$}
\def \triax {$\overline{T}$}
\def \sigellip {$\sigma_{E}$}
\def \sigtriax {$\sigma_{T}$}
\def \galfit {{\sc galfit}}
\def \ellipse {{\sc ellipse}}
\def \iraf {{\sc iraf}}
\def \emcee {{\sc emcee}}
\def \mstar {$M_{*}$}
\def \msun {$M_{\odot}$}
\def \lsun {$L_{\odot}$}
\def \atlas {ATLAS$^{3\text{D}}$}
\def \reffig {Fig.\,{}}

\newcommand{\rsj}[1] {\textcolor{blue}{(rsj: #1)}}
\newcommand{\new}[1] {\textcolor{black}{#1}}
\newcommand{\vect}[1]{\boldsymbol{\mathbf{#1}}}
\newcommand{\myfig}[1]{Fig.\,\ref{#1}}

\newcommand{\Engvs}{0.43}
\newcommand{\Eungvs}{+0.02}
\newcommand{\Elngvs}{-0.02}


\title{The Next Generation Virgo Cluster Survey. XXIII. Fundamentals of nuclear star clusters over seven decades in galaxy mass}

\shorttitle{NSCs in Virgo galaxies}
\shortauthors{S\'anchez-Janssen et al.}


\author{
Rub\'en S\'anchez-Janssen\altaffilmark{\ref{ukatc},\ref{nrc}}, %
Patrick C\^ot\'e\altaffilmark{\ref{nrc}}, %
Laura Ferrarese\altaffilmark{\ref{nrc},\ref{gem-n}}, %
Eric W. Peng\altaffilmark{\ref{peking},\ref{kavli}}, %
Joel Roediger\altaffilmark{\ref{nrc}}, %
John P. Blakeslee\altaffilmark{\ref{nrc}}, %
Eric Emsellem\altaffilmark{\ref{eso},\ref{lyon}}, %
Thomas H. Puzia\altaffilmark{\ref{puc}}, %
Chelsea Spengler\altaffilmark{\ref{nrc}}, %
James Taylor\altaffilmark{\ref{waterloo}},
Karla A. \'Alamo-Mart\'inez\altaffilmark{\ref{puc}}, %
Alessandro Boselli\altaffilmark{\ref{lam}}, %
Michele Cantiello\altaffilmark{\ref{teramo}}, %
Jean-Charles Cuillandre\altaffilmark{\ref{cfht}}, %
Pierre-Alain Duc\altaffilmark{\ref{stras},\ref{cea}}, %
Patrick Durrell\altaffilmark{\ref{ohio}}, %
Stephen Gwyn\altaffilmark{\ref{nrc}}, %
Lauren A. MacArthur\altaffilmark{\ref{princeton}}, %
Ariane Lan\c{c}on\altaffilmark{\ref{stras}}, %
Sungsoon Lim\altaffilmark{\ref{peking},\ref{kavli}},
Chengze Liu\altaffilmark{\ref{shanghai}}, %
Simona Mei\altaffilmark{\ref{gepi},\ref{paris},\ref{caltech}}, %
Bryan Miller\altaffilmark{\ref{gemini}}, %
Roberto Mu\~noz\altaffilmark{\ref{puc}}, %
J. Christopher Mihos\altaffilmark{\ref{case}}, %
Sanjaya Paudel\altaffilmark{\ref{korea}}, %
Mathieu Powalka\altaffilmark{\ref{stras}}, %
Elisa Toloba\altaffilmark{\ref{pacific}}%
}

\email{ruben.sanchez-janssen@stfc.ac.uk}

\newcounter{address}
\setcounter{address}{1}
\altaffiltext{\theaddress}{\refstepcounter{address}\label{ukatc}STFC UK Astronomy Technology Centre,  Royal Observatory, Blackford Hill, Edinburgh, EH9 3HJ, UK}
\altaffiltext{\theaddress}{\refstepcounter{address}\label{nrc}NRC Herzberg Astronomy and Astrophysics, 5071 West Saanich Road, Victoria, BC, V9E 2E7, Canada}
\altaffiltext{\theaddress}{\refstepcounter{address}\label{gem-n} Gemini Observatory, Northern Operations Center, 670 N. A?ohoku Place, Hilo, HI 96720, USA}
\altaffiltext{\theaddress}{\refstepcounter{address}\label{peking} Department of Astronomy, Peking University, Beijing 100871, China}
\altaffiltext{\theaddress}{\refstepcounter{address}\label{kavli} Kavli Institute for Astronomy and Astrophysics, Peking University, Beijing 100871, China}
\altaffiltext{\theaddress}{\refstepcounter{address}\label{eso} European Southern Observatory, Karl-Schwarzschild-Str. 2, 85748 Garching, Germany}
\altaffiltext{\theaddress}{\refstepcounter{address}\label{lyon} Centre de Recherche Astrophysique de Lyon and \'Ecole Normale Sup\'erieure de Lyon, Observatoire de Lyon, Université Lyon 1, 9 avenue Charles André, F-69230 Saint-Genis Laval, France}
\altaffiltext{\theaddress}{\refstepcounter{address}\label{puc} Institute of Astrophysics, Pontificia Universidad Cat\'olica de Chile, Av. Vicu\~na Mackenna 4860, 7820436 Macul, Santiago, Chile}
\altaffiltext{\theaddress}{\refstepcounter{address}\label{waterloo}Department of Physics and Astronomy, University of Waterloo, Waterloo, ON N2L 3G1, Canada}
\altaffiltext{\theaddress}{\refstepcounter{address}\label{lam} Laboratoire d'Astrophysique de Marseille - LAM, Universit\'e d'Aix-Marseille \& CNRS, UMR7326, 38 rue F. Joliot-Curie, 13388, Marseille Cedex 13, France}
\altaffiltext{\theaddress}{\refstepcounter{address}\label{teramo} INAF-Osservatorio Astronomico di Teramo, Italy}
\altaffiltext{\theaddress}{\refstepcounter{address}\label{cfht} Canada--France--Hawaii Telescope Corporation, Kamuela, HI 96743, USA}
\altaffiltext{\theaddress}{\refstepcounter{address}\label{stras}Observatoire Astronomique, Université de Strasbourg \& CNRS UMR 7550, 11 rue de l?Université, 67000 Strasbourg, France}
\altaffiltext{\theaddress}{\refstepcounter{address}\label{cea} AIM Paris Saclay, CNRS/INSU, CEA/Irfu, Universit\'e Paris Diderot, Orme des Merisiers, F-91191 Gif sur Yvette cedex, France}
\altaffiltext{\theaddress}{\refstepcounter{address}\label{ohio} Department of Physics and Astronomy, Youngstown State University, One University Plaza, Youngstown, OH 44555, USA}
\altaffiltext{\theaddress}{\refstepcounter{address}\label{princeton} Department of Astrophysical Sciences, Princeton University, Princeton, NJ 08544, USA}
\altaffiltext{\theaddress}{\refstepcounter{address}\label{shanghai} Center for Astronomy and Astrophysics, Department of Physics and Astronomy, Shanghai Jiao Tong University, Shanghai 200240, China}
\altaffiltext{\theaddress}{\refstepcounter{address}\label{gepi} LERMA, Observatoire de Paris,  PSL Research University, CNRS, Sorbonne Universit\'es, UPMC Univ. Paris 06, F-75014 Paris, France}
\altaffiltext{\theaddress}{\refstepcounter{address}\label{paris} University of Paris Denis Diderot, University of Paris Sorbonne Cit\'e (PSC), 75205 Paris Cedex 13, France}
\altaffiltext{\theaddress}{\refstepcounter{address}\label{caltech} Jet Propulsion Laboratory, Cahill Center for Astronomy \& Astrophysics, California Institute of Technology, 4800 Oak Grove Drive, Pasadena, California, USA}
\altaffiltext{\theaddress}{\refstepcounter{address}\label{gemini} Gemini Observatory, Casilla 603, La Serena, Chile}
\altaffiltext{\theaddress}{\refstepcounter{address}\label{case}Department of Astronomy, Case Western Reserve University, Cleveland, OH, USA}
\altaffiltext{\theaddress}{\refstepcounter{address}\label{korea} Korea Astronomy and Space Science Institute, Daejeon 305-348, Republic of Korea}
\altaffiltext{\theaddress}{\refstepcounter{address}\label{pacific} Department of Physics, University of the Pacific, 3601 Pacific Avenue, Stockton, CA 95211, USA}

\begin{abstract}
Using deep, high resolution optical imaging from the Next Generation Virgo Cluster Survey we study the properties of nuclear star clusters (NSCs) in a sample of nearly 400 quiescent galaxies in the core of Virgo with stellar masses $10^{5}\lesssim$\mstar/\msun$\lesssim10^{12}$. 
The nucleation fraction reaches a peak value $f_{n}\approx90\%$ for \mstar~$\approx10^{9}$\msun\ galaxies and declines for both higher and lower  masses, but nuclei populate galaxies as small as \mstar~$\approx5\times10^{5}$\msun. 
Comparison with literature data for nearby groups and clusters shows that at the low-mass end nucleation is more frequent in denser environments. 
The NSC mass function peaks at $M_{NSC}\approx7\times10^{5}$\msun, a factor 3-4 times larger than the turnover mass for globular clusters (GCs). 
We find a  nonlinear  relation between the stellar masses of NSCs and of their host galaxies, with a mean nucleus-to-galaxy mass ratio that drops to $M_{NSC}/M_{\star}\approx3.6\times10^{-3}$ for \mstar~$\approx5\times10^{9}$\msun\ galaxies. 
Nuclei in both more  and less massive galaxies are much more prominent: $M_{NSC}\propto M_{*}^{0.46}$ at the low-mass end, where nuclei are nearly 50\% as massive as their hosts. 
We measure an intrinsic scatter in NSC masses at fixed galaxy stellar mass of 0.4\,dex, which we interpret as evidence that the process of NSC growth is significantly stochastic. 
At low galaxy masses we find a close connection between NSCs and GC systems, including a very similar occupation distribution and comparable total masses.
We discuss these results in the context of current dissipative and dissipationless models of NSC formation.  
\end{abstract}



\section{Introduction}
\label{sect:intro}


The very central regions of galaxies are extreme astrophysical  environments, sites where numerous complex mechanisms operate simultaneously.
They mark the bottom of the galactic potential well, where matter has been accumulating throughout the entire galaxy history.
A fraction of their constituent material can be traced back to the earliest generations of stars  that formed during the rapid collapse of the rarest density peaks \citep{Diemand2005}, but also to more recent  epochs as a product of recurrent gas inflows followed by associated star formation events.
As a result, these inner regions feature the highest  galactic stellar densities  and the shortest relaxation times, and it is no accident that they also harbor the two types of known compact massive objects (CMOs), namely massive black holes (MBHs) and nuclear star clusters (NSCs). 

NSCs are compact stellar systems with half-light radii in the range of 1-50 pc and stellar masses stretching from as low as $10^{4}$ \msun\ to as high as $10^{8}$ \msun. 
On average, they tend to be larger and more massive than the typical GC but, remarkably, their central stellar surface densities can be even more extreme--amongst the highest known, sometimes exceeding $\Sigma_{c} = 10^{5}$ \msun\,pc$^{-2}$ \citep[e.g.,][]{Lauer1998,Hopkins2010a}.
These central stellar densities are only rivalled by some GCs and ultra-compact dwarfs (UCDs) \citep[][]{Hilker1999,Drinkwater2000}.
UCDs also present remarkable similarities with NSCs in terms of their stellar population content \citep{Chilingarian2008,Norris2014,Liu2015b}.
All this, combined with the fact that UCDs tend to live in close proximity to massive galaxies \citep{Hasegan2005,Hau2009,Liu2015b} and as a population have distinct kinematical properties from the galactic GC systems \citep{Zhang2015} has led to the suggestion that a non-negligible fraction of UCDs may represent the remnants of tidally disrupted nucleated galaxies \citep[][but see \citealt{Mieske2006} and \citealt{DePropris2005} for arguments against this scenario]{Drinkwater2003,Pfeffer2013}.
The discovery \citep{Seth2014,Ahn2017} that M60-UCD1 in the Virgo cluster harbors a MBH contributing $\approx$\,15\% of the total mass lends further support to the nuclear origin of (at least) the most massive UCDs.
The threshing scenario has also been put forward to explain the chemical, structural and dynamical anomalies of the most massive GCs in the Local Group, like $\omega$Cen in the  Galaxy \citep{Lee1999,Hilker2000,Bekki2003a} and Mayall\,II/G1 in M31 \citep{Meylan2001,Bekki2004,Ma2007}. 

NSCs inhabit galaxies spanning a wide range of masses, morphological types and gas content, and nucleation seems to be a complex function of all these parameters \citep{Binggeli1987,Binggeli1991,Carollo1998,Boker2002,Walcher2005,Cote2006,Cote2007,Seth2006,Lisker2007,Georgiev2009b,Glass2011,Turner2012,denBrok2014,Georgiev2014}. 
Interestingly, galaxies and NSCs display a variety of scaling relations, including with their stellar masses \citep[e.g.,][]{Scott2013,Georgiev2016} and their stellar populations or colors \citep{Walcher2005,Turner2012,Georgiev2014}.
This suggests that their formation is intricately linked to that of their host galaxy. 
And perhaps not entirely surprising, some of these scaling relations may be similar to those followed by MBHs, a picture that further relates the two families of CMOs \citep{Ferrarese2006a,Wehner2006,Graham2009}. 

Formation scenarios for NSCs can be broadly divided in two categories.
The first one involves a dissipationless process, whereby the orbits of pre-existing dense star clusters decay as a result of dynamical friction and produce mergers in the central regions of the galaxy \citep{Tremaine1975, Capuzzo-Dolcetta1993, Capuzzo-Dolcetta2008a, Capuzzo-Dolcetta2008, Agarwal2011,Gnedin2014,Arca-Sedda2014}.
The second scenario consists of a dissipative mode through in situ central star formation driven by gas inflows and condensation \citep{Bekki2006c,Bekki2007,Antonini2015}, with the first seeds developing perhaps as early as the epoch of reionization \citep{Cen2001}. 
The latter growth mechanism is of course regulated by the availability of sufficient gas reservoirs in the central regions and internal feedback mechanisms. 
Most likely both processes contribute to some degree to the formation of the NSCs we observe in the Local Universe \citep{Hartmann2011, Antonini2015}, and hybrid scenarios involving the coalescence of gas-rich star clusters have been proposed \citep{Guillard2016}. 
The heterogeneity of the proposed formation scenarios is consistent with observations of the stellar populations in NSCs. 
They tend to be rather complex, showing evidence for multiple generation of stars, and younger mean ages and higher metallicities than typical in GCs \citep{Rossa2006,Walcher2006,Puzia2008,Paudel2011}.
But the relative weight of the different formation mechanisms, and how this breakdown depends on host galaxy properties, are unknown.
Related, and perhaps more fundamental questions are what determines whether a galaxy forms an NSC in the first place, and what mechanisms regulate their subsequent growth. 

Considering all the unknowns about the origin and growth of NSCs it is fundamental to accurately characterize their occurrence and how they relate to their host galaxies.
In this contribution we take a step forward in this direction and present a comprehensive study on the abundance and properties of NSCs in a volume- and mass-limited sample of galaxies in the Virgo cluster spanning seven decades in stellar mass. 
Focusing our study on  cluster galaxies presents several advantages. 
First, the high densities in clusters directly translate into large galaxy samples and in studies with significant statistical power. 
Second, it is also straightforward to translate abundance results into volume-limited quantities--and this is a much harder exercise in the field. 
Finally, because all these galaxies share a common (overdense) environment, statistically speaking they first collapsed at similarly early epochs and have been subject to roughly the same amount of environmental effects.
\new{In the particular case of quiescent galaxies in the cores of clusters as massive as Virgo the star formation activity is expected to have  ceased at least 5-6 billion years ago \citep[e.g.,][]{Mistani2016}. 
The lack of substantial star formation activity in the recent past for these red-sequence cluster galaxies implies that their NSCs should have been polluted to a lesser degree than those at the centers of gas-rich, more luminous disc galaxies--but our knowledge about the stellar populations of faint NSCs is still limited \citep{Spengler2017,Ordenes-Briceno2018}.}
Readers interested in the properties of NSCs in  those star forming systems  are referred to the recent works by \citet{Georgiev2014}, \citet{Carson2015} and \citet{Georgiev2016}. \citet{Georgiev2009b} present a study on the nuclei of lower mass dwarf irregulars.

This paper is organized as follows.
Section\,\ref{sect:data} describes the methods for NSC detection and the measurement of their properties in the NGVS images.
In Section\,\ref{sect:fraction} we study the nucleation fraction in the core of the Virgo cluster, followed by an investigation of the scaling relations between NSCs and their host galaxies in Section\,\ref{sect:scalings}.   
Section\,\ref{sect:discussion} presents a discussion on the previous results in the context of NSC formation models and on the relationship to other star cluster systems.
Finally, in Section\,\ref{sect:conclusions} we summarize the main findings of this work, and lay out our conclusions.
Throughout this manuscript we use a common distance modulus $(m-M)=31.09$ mag for all candidate Virgo members, corresponding to the mean distance of $D=16.5$ Mpc to the Virgo cluster derived through the surface brightness fluctuations method \citep{Mei2007,Blakeslee2009}. This translates into a physical scale of 80 pc per arcsecond.

\section{Detection and characterization of NSCs in the NGVS}
\label{sect:data}

Detailed descriptions of the NGVS and  its associated data products are given elsewhere \citep{Ferrarese2012,Ferrarese2016}. 
Briefly, here we only use data in the core of Virgo, which for our purposes refers to the square region roughly centered on M87 and 2 deg = 0.58 Mpc = 0.37\,$R_{vir}$ on the side.
This area was imaged in the  $u'griz'K_{s}$ bandpasses, and galaxies were detected using a ring median filter algorithm optimized to extract low surface brightness objects.
Virgo members were then identified using a combination of colors and structural and quantitative morphological parameters, which were further complemented with a visual inspection by several members of our team.
This process resulted in a parent sample of 404 galaxies in the core of Virgo spanning approximately seven decades in stellar mass, $10^{5} \lesssim M_{*}/M_{\odot} \lesssim 10^{12}$ \citep{Ferrarese2016}.
In this study we are only interested in the NSCs of quiescent galaxies, and we accordingly discard 24 objects that have evidence for ongoing star formation activity \citep[see][]{Roediger2016}.

\begin{figure}[!t]
\includegraphics[angle=0,width=0.5\textwidth]{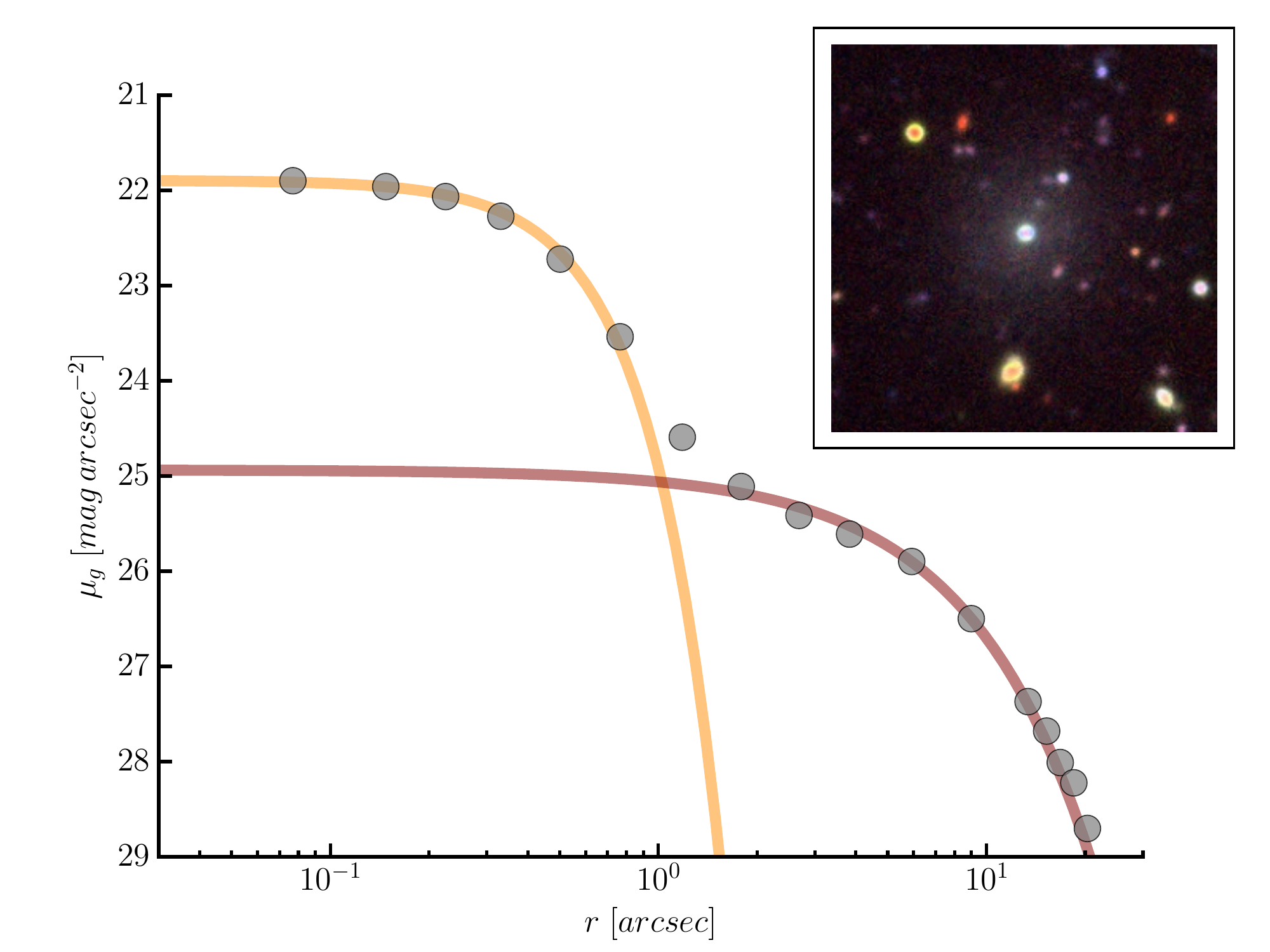}
\caption{Structural characterization of the nucleated low mass galaxy VCC\,1070 in the NGVS.
The grey points show the surface brightness profile as measured by \ellipse. 
The maroon and orange lines  correspond to the seeing-convolved best-fit Sersic profiles to the galaxy and NSC components, respectively. 
The inset image shows a $gri$ composite of the galaxy. Note the remarkable prominence of the NSC in VCC\,1070.}
\label{fig:sbp}
\end{figure}

\new{The structural characterization of these galaxies is presented in full  in Ferrarese et al. (2018). 
Here we only provide the most salient details pertaining to the analysis of NSCs.
A full isophotal analysis is carried out using a semi-automated code that (i) extracts image cutouts and masks contaminants and cosmetics; (ii) performs isophotal fitting and extracts one-dimensional profiles for the surface brightness, ellipticity, major axis position angle, isophotal centre, and deviations of the isophotes from pure ellipses; (iii) carries out parametric fits to the surface brightness profiles while accounting for the effects of the PSF (see \myfig{fig:sbp}).
The galaxy body is modeled with a single \citet{Sersic1968} function, which provides an adequate description of these smooth, nearly oblate galaxies \citep{Sanchez-Janssen2016}.
If necessary, a central nuclear component is included in the fits. 
NSCs are also modeled with S\'ersic functions, but with a few exceptions these nuclei remain unresolved despite the exquisite image quality of the NGVS (PSF FWHM $\approx 0.6$\,arcsec in the $i$-band).
All the quoted errors correspond to the one-sigma formal uncertainty of the fitted S\'ersic profiles.
The analysis is carried out independently in each of the photometric bands. 
}

\new{The detection of barely resolved or unresolved NSCs is always challenging. 
The operational definition of NSCs in this paper requires the existence of a luminosity excess above the main stellar distribution in the core regions of the galaxies. 
The identification of such objects is based on the relative quality of the S\'ersic and double-S\'ersic fits (as measured by the fit $\chi^{2}$), complemented by a visual inspection of color and unsharp-masked images of the galaxies.
In the Virgo core we classify 107 galaxies as nucleated.
The formal limiting magnitude for unresolved sources in the NGVS is $g = 25.9$\,mag \citep{Ferrarese2012}, and this translates into a limiting mass for NSCs log\,($M_{NSC}/$\msun)~$\approx4.5$. 
However, because the nuclei are additionally visually classified, the effective detection threshold in this study may be slightly higher (see details in Sect.\,\ref{sect:scaling}).
}

Stellar masses for the galaxies and the NSCs are obtained through modeling  of their spectral energy distributions (SEDs) in the $u'griz'$ bands. 
Details will be presented in Roediger et al. (in prep.), but essentially we employ the Flexible Stellar Population Synthesis models of \citep{Conroy2009a}, assuming exponentially declining star formation histories and a Chabrier initial mass function. The SEDs are fit to a  grid of 50,000 synthetic models with metalliticies in the  $0.01 \leq Z/Z_{\odot} \leq 1.6$ range, star formation timescales  $0.5  \leq \tau \leq 100$ Gyr$^{-1}$, and luminosity-weighted ages between 5 and 13 Gyr.
In this work we carry out several comparisons between the NGVS data and other samples from the literature for which  multiwavelength photometry is not always available.
Thus for the literature samples we convert the luminosities to masses using the \mstar/$L$ ratios derived from the above SED fitting procedure.
Specifically, for galaxies we use the median relation between the \mstar/$L$ ratios as a function of luminosity in the corresponding band (usually $g$ or $i$), so that we account for mass-dependent variations.
For NSCs we simply use the median \mstar/$L$ from our sample of nuclei. 
Typical uncertainties for these stellar masses are $\sigma({log\,M_{*}}) \approx 0.15$ dex and $\sigma({log\,M_{NSC}}) \approx 0.20$ dex.

Because faint NSCs are unresolved and hard to distinguish from stars we have quantified the likelihood of contamination by chance superposition of Galactic stellar interlopers over the geometric center of a galaxy as follows. 
We first computed  the density of stars in the Virgo core, $\rho_{\star}$, using a catalog of largely unresolved sources brighter than $g=25$ mag (0.2\,mag fainter than the faintest NSC in the sample). 
These sources have been assigned probabilities to be either stars, GCs or background galaxies using a mixture model algorithm that incorporates information about their spectral energy distribution and structure (Peng et al., in preparation). 
\new{From that catalog we only select the $\approx$\,5500 sources with $P(\text{star}) > 0.5$ that have colors $0.5 < (g-i) < 1.15$, which is roughly the range expected for NSCs and UCDs \citep[cf.][and \myfig{fig:colors}]{Liu2015b}.
We further assume a maximum nucleus-galaxy offset of $R = 2$ arcsec (equivalent to 160 pc), which is the maximum distance found by \citet{Cote2006}) in their analysis of ACSVCS nuclei. In our case the great majority of nuclei are found within 0.4 arcsec of the galaxy centre (Ferrarese et al. 2018). 
Using these numbers we estimate a total number of $\rho_{\star}\pi R^{2} \approx 1.5\times10^{-3}$ stars per galaxy. 
Equivalently, this implies that in this worst-case scenario no more than one galaxy in the Virgo core would be affected by a star being close enough to its center to be mistaken for an NSC.
}

\section{Nucleation fraction}
\label{sect:fraction}

\begin{figure}[!t]
\includegraphics[angle=0,width=0.5\textwidth]{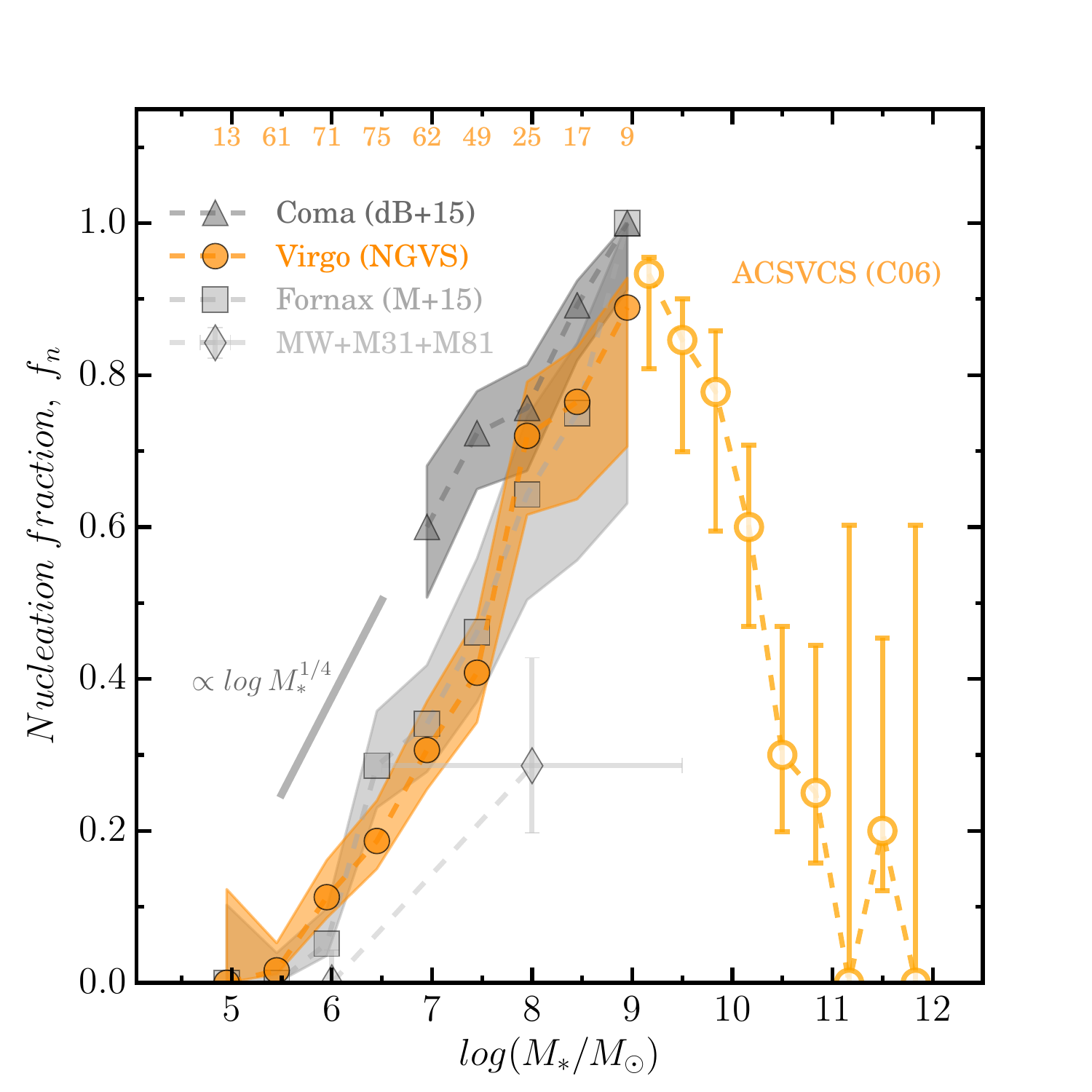}
\caption{Fraction of nucleated galaxies  in different environments as a function of galaxy stellar mass.  The Virgo, Fornax, and Coma clusters are represented with circles, squares and triangles, respectively. The two diamond symbols correspond to the satellites of  three nearby spiral-dominated galaxy groups. Shaded regions and error bars indicate the corresponding 68 per cent Bayesian credible interval, and the figures at the top show the total number of NGVS galaxies in each stellar mass bin. Note how, at fixed stellar mass, the fraction of nucleated galaxies in Virgo is always lower than in Coma. The rapid decline toward low masses implies that the faintest nucleated low-mass galaxy in Virgo has a stellar mass \mstar~$\approx 5\times10^{5}$~\msun.}
\label{fig:frac}
\end{figure}

We first turn our attention to the  occurrence of NSCs in the core of Virgo. 
In \myfig{fig:frac} we show with solid circles $f_{n}$, the fraction of nucleated galaxies in the NGVS as a function of galaxy stellar mass for objects with masses \mstar~$\leq10^{9}$ \msun. 
The amount of galaxies in the Virgo core area more massive than this limit drops quickly, as indicated by the numbers at the top of \myfig{fig:frac}.
This results in too uncertain estimates of the nucleation fraction for more luminous galaxies.
For completeness, in the  $10^{9} < M_{*}/M_{\odot}< 10^{12}$ mass range we derive the fraction of nucleated galaxies in Virgo from the ACSVCS \citep{Cote2006}, which is not limited to the core region and therefore has stronger statistical power at the high mass end. 
Throughout this paper, and unless otherwise stated, all uncertainties associated to binomial proportions correspond to the 68\% Bayesian credible interval. 

Consistent with previous work we find that the nucleation fraction is very high at intermediate masses, such that at  \mstar~$\approx 10^{9}$ \msun\ over $90\%$ of the galaxies in the core of Virgo harbor an NSC.
The nucleation fraction was found by \citet{Cote2006} to drop sharply for massive galaxies, a behavior that is often attributed to the highly disruptive power of the MBHs that inhabit the central regions of these galaxies (see \citealt{Antonini2013}, and Sect.\,\ref{sect:discussion}). 
Interestingly, we find that $f_{n}$ also decreases toward lower masses, albeit at a slightly slower rate so that the NSC occupation distribution almost resembles a lognormal function. 
At low masses this decline is well  described by $f_{n} \propto \mbox{log\,\mstar}^{1/4}$. 
We identify a threshold value below which no low-mass galaxy with $\langle\mu_{g}\rangle_{e} < 28$ \sb\ in the core of Virgo is nucleated, \mstar\ $\approx 5\times10^{5}$ \msun\ (see \myfig{fig:frac}). 
This indicates that either NSC formation is highly inefficient in low-mass halos or, alternatively, that some mechanism enhances NSC disruption in these shallow potentials. 
The strong dependence of the nucleation fraction on galaxy mass also implies that care must be taken when comparing different samples if they span different mass ranges. 
The \emph{global} nucleation fraction is an ill-defined quantity, unless the mass range under study is specified.

In \myfig{fig:frac} we also explore the dependence of the nucleation fraction on global environment, that is, the mass of the host potential where the galaxies reside. 
We include data on the fraction of early-type nucleated galaxies in the Coma \citep{denBrok2014} and Fornax clusters \citep{Munoz2015}, as well in three spiral-dominated groups in the Local Universe, namely the Milky Way (MW), M31 and M81. 
\new{For these two datasets we use the published $i$-band magnitudes and luminosity-dependent mass-to-light ratios that vary non-linearly in the range $0.5 < M_{*}/L_{i} < 1.7$ \citep[cf.][]{Zhang2017}.
For the NSCs we use the median $M_{*}/L_{i} =1.3$ from our sample of Virgo nuclei.}
Details of the methodology used to identify nucleated galaxies  in the nearby groups and to estimate their stellar masses are provided in Appendix\,\ref{sect:appendix}. 
It is also important to remark that the effective spatial resolution is almost identical for the different cluster samples, because the larger distance of Coma is compensated by the use of HST/ACS imaging.
We thus sample a wide range in host halo masses, from $M_{h} \approx 10^{15}$ \msun\ for Coma, $M_{h} \approx 10^{14}$ \msun\ for Virgo and Fornax, and $M_{h} \approx 10^{12}$ \msun\ for the three groups with $L^{*}$ centrals. 
We note that the three cluster samples only include galaxies from roughly the same physical regions, 0.2-0.25\,$R_{\text{vir}}$, and so can be directly compared. 

We find that in all environments the nucleation fraction is a similarly strong function of galaxy stellar mass, but we unveil a secondary dependence on host halo mass: at fixed \mstar\ the fraction of galaxies harboring NSCs is larger in denser environments. 
While the $f_{n}$ curves in Virgo and Fornax display an almost identical behavior, at  all stellar masses the nucleation fraction  is systematically larger  in Coma and lower in the nearby groups. 
The effect is perhaps best appreciated by comparing the integrated nucleation fraction within the common  [$10^{7},10^{9}$] \msun\ mass range, where 77\%, 53\% and 56\% of the galaxies host an NSC in Coma, Virgo, and Fornax, respectively. 
The figure for the three spiral-dominated groups combined drops to 29\%, but it is necessarily a noisier measurement due to the reduced satellite sample size (only 14 galaxies in this mass range). 
It would be interesting to explore to which extent the high nucleation fraction in massive clusters like Coma holds toward lower stellar masses, i.e., whether the threshold mass for NSC occurrence we find in Virgo is roughly universal, or also varies with host halo mass.   
We conclude that while stellar mass is the main parameter governing NSC occurrence, their abundance is also enhanced in high-density environments.


\section{Scaling relations between NSCs and their host galaxies}
\label{sect:scalings}

Once we have established the frequency with which NSCs occur in Virgo galaxies, we explore their colors, their masses, and the relation to the properties of their host galaxies.

\subsection{Colors of NSCs}

\begin{figure}[!t]
\includegraphics[angle=0,width=0.5\textwidth]{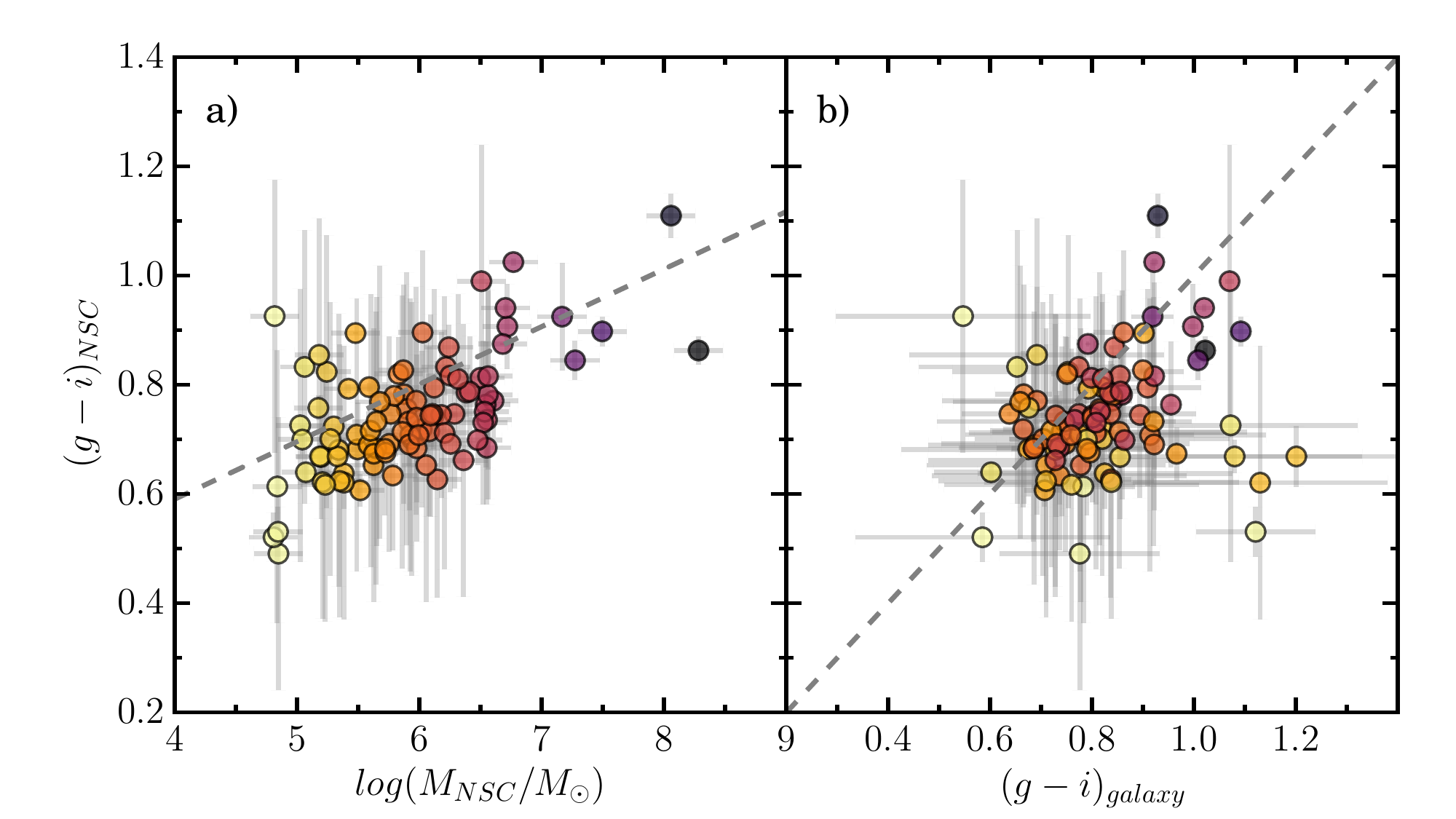}
\caption{The colors of NSCs in the core of the Virgo cluster. 
{\it Left}: the $(g-i)$ colors of NSCs are shown against the mass of the nucleus. The two quantities are correlated, and the dashed line shows the best-fit relation. 
{\it Right}: the colors are now plotted against the color of the host galaxy. The correlation is much weaker. 
In both panels the symbols are color-coded according to $M_{NSC}$.}
\label{fig:colors}
\end{figure}

In \myfig{fig:colors}a we show the relation between NSC masses and their $(g-i)$ colors for the galaxies in the core of Virgo. 
We note that this is a slightly reduced subsample where we have excluded 12 objects that have highly uncertain $(g-i)$ colors ($\sigma_{c} > 0.3$ mag).
Consistent with previous work \citep[e.g.,][]{Turner2012} we find that the colors of NSCs correlate with $M_{NSC}$, such that more massive nuclei tend to be slightly redder (Pearson's correlation coefficient $r = 0.6$). 
The dashed line in this panel shows the best-fit relation, $(g-i) = 0.17 + 0.1\,\text{log}\,M_{NSC}$.
The median color for the sample is $(g-i) = 0.73$, which as shown by \citet{Roediger2016} is remarkably consistent with the peak of the galaxy color distribution at low masses.  
In the right-hand panel,  \myfig{fig:colors}b shows the same colors now plotted against the colors of the host galaxies. 
The dashed line indicates the identity relation. 
The two quantities are only weakly correlated ($r = 0.23$), and we consider this to be an indication that the connection is only a secondary effect.
As we will show in the next section, less massive galaxies (which are bluer) tend to have less massive NSCs. 
And because the mass of the NSC is correlated with its color, we naturally find marginally bluer nuclei in bluer galaxies.
\new{The mean color difference is only $\langle\Delta(g-i)\rangle = \langle (g-i)_{galaxy} - (g-i)_{NSC} \rangle = 0.06$, which indicates that NSCs have marginally bluer colors than their host galaxies.
This is ratified upon inspection of the fraction of NSCs that have $\langle\Delta(g-i)\rangle > 0$, which amounts to 75\%. 
}
This is in agreement with previous work \citep{Paudel2011}, indicating that NSCs may be marginally younger and/or more metal poor than the bulk of stars in the galactic body.
A more detailed analysis of the SEDs and the stellar population content of NSCs using NGVS data is presented in \citet{Spengler2017}.

\subsection{Relation to the host galaxy}

\subsubsection{The NSC-to-galaxy mass relation}
\label{sect:scaling}

\begin{figure}[!t]
\includegraphics[angle=0,width=0.5\textwidth]{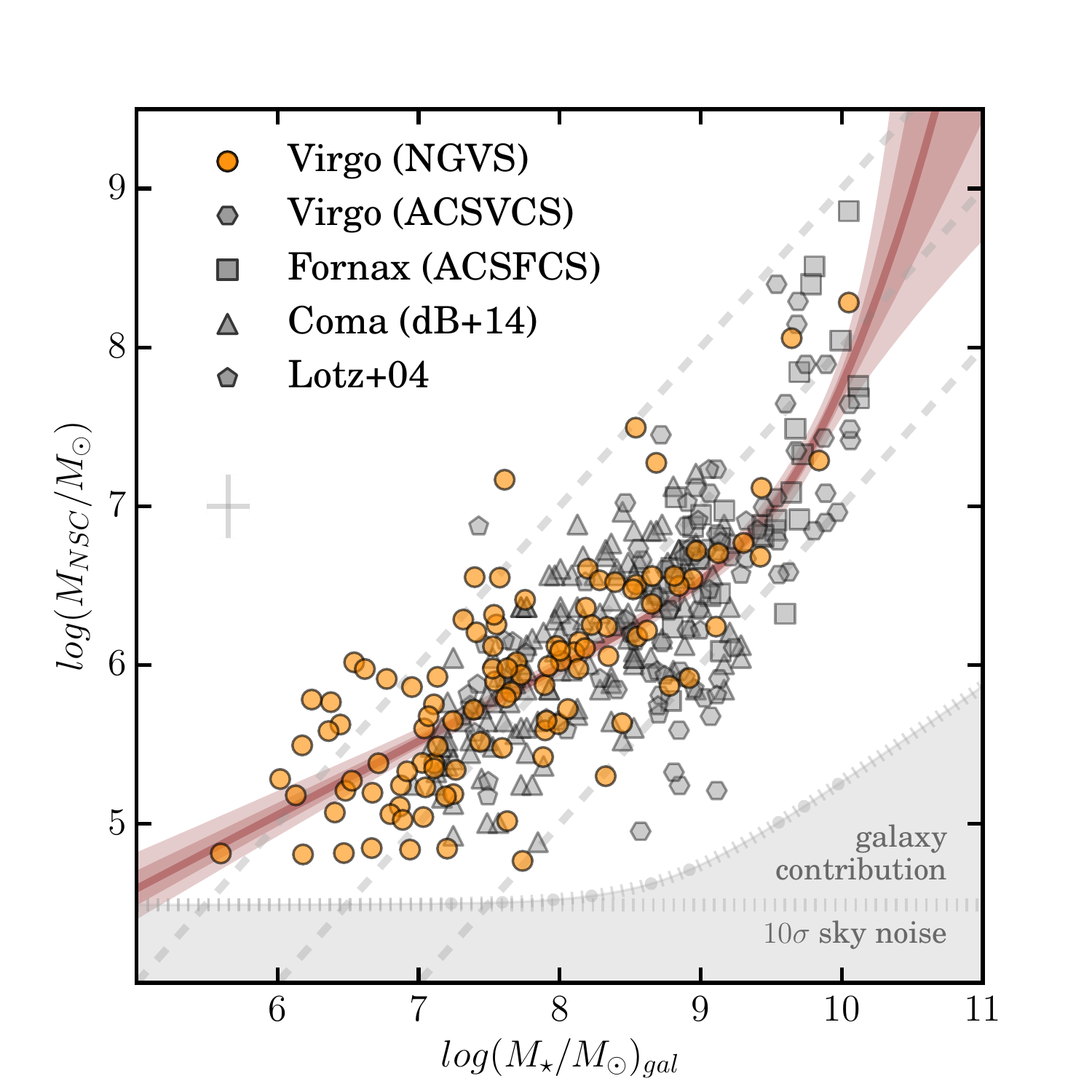}
\caption{The NSC-to-galaxy mass relation. 
The NGVS data for the core region is represented with circles, but here we also include literature data for early-type galaxies in Virgo, Fornax, and Coma (see text for details). 
The error bar shows the typical uncertainty in  the mass estimates. 
The shaded region at the bottom indicates the NSC mass equivalent to the $10\sigma$ detection limit in the NGVS, including the contribution of the  underlying galaxy light.
The three dashed lines, from right to left, correspond to constant NSC mass fractions of 0.1\%, 1\% and 10\%.
Note that while more massive galaxies harbor more massive nuclei, the relation is nonlinear. 
The best-fit relation is plotted as a solid line together with the 68\% and 95\% confidence intervals. We measure an intrinsic scatter for the relation of nearly 0.4 dex, which we interpret as evidence for stochastic NSC growth.}
\label{fig:scaling}
\end{figure}

In  \myfig{fig:scaling} we show the relation between NSC stellar masses and those of their host galaxies.
The NGVS data for the core region is represented with circles, but here we also include data for other early-type galaxies in Virgo (from the ACSVCS \citealt{Cote2004}), Fornax (ACSFCS, \citealt{Turner2012}), Coma \citep{denBrok2014}, and the sample of Virgo and Fornax faint dEs from \citet{Lotz2004}. 
The three dashed lines, from right to left, correspond to constant NSC mass fractions of 0.1\%, 1\% and 10\%.
The shaded region at the bottom of the panel indicates the NSC mass equivalent to the $10\sigma$ detection limit for unresolved sources in the NGVS, log\,($M_{NSC}/$\msun)~$\approx4.5$ (see \citealt{Ferrarese2012}). 
Two sources contribute to the local background against which NSCs are detected, namely the underlying galaxy stellar light, and the sky brightness. 
Here the galaxy term is estimated by computing the Poisson noise associated to the average galaxy central surface brightness as a function of stellar mass. It is only dominant for relatively massive galaxies, log\,($M_{NSC}/$\msun)~$\gtrsim 8.5$, whereas the sky brightness is the main source of noise for fainter systems.
\new{We note that our effective detection threshold may be slightly higher than shown. NSCs are additionally visually classified, and it is possible that in this process we have discarded extremely faint nuclei that may fall below the detection limit in a single band (therefore giving it a suspicious visual appearance).}

An important conclusion that can be readily drawn from  \myfig{fig:scaling} is that, contrary to what we have found regarding the nucleation fraction, the NSC-to-galaxy stellar mass relation for early-type cluster galaxies seems to be independent of environment. All galaxies in Virgo, Fornax and Coma exhibit the same behavior, albeit with large intrinsic scatter. 
The universality of the NSC-to-galaxy stellar mass relation suggests that the mass of the NSC is primarily controlled by the galaxy mass, but its large intrinsic scatter indicates that the process of mass deposition in the nuclear region probably is quite stochastic in nature (see discussion below).
It is also evident that even though more massive galaxies host the most massive nuclei the relation is  nonlinear, featuring a significant steepening at the massive end.
The shape of the relation is very reminiscent of the stellar-to-halo mass relation (SHMR), so we provide a quantitive description using the following 5-parameter functional form:

\begin{equation}
\begin{aligned}
\mbox{log}\,M_{NSC}~=~& \mbox{log}\,M^{'}_{NSC}  ~-~ \frac{1}{2} \\
				       & ~+~  \beta\,\mbox{log}\bigg(\frac{M_{*}}{M^{'}_{*}}\bigg) \\
				       & ~+~ \frac{\big(\frac{M_{*}}{M^{'}_{*}}\big)^{\delta}}{1 + \big(\frac{M_{*}}{M^{'}_{*}}\big)^{-\gamma}}, 
\end{aligned}
\label{eq:fit}
\end{equation}

\noindent where $M^{'}_{NSC}$ is a characteristic NSC mass, $M^{'}_{*}$ is a characteristic galaxy stellar mass, $\beta$ is the slope of the relation at the low mass end, and $\delta$ and $\gamma$ determine the massive end slope.
This expression has been shown to provide an adequate description of the SHMR in all environments and at all redshifts \citep[see ][]{Behroozi2013,Grossauer2015}, and will help us better interpret our results on NSC properties in the larger context of galaxy formation in a $\Lambda$CDM framework.

We carry out a Bayesian fit to Eq.\,\ref{eq:fit} while simultaneously allowing for intrinsic scatter at fixed galaxy stellar mass, parametrized by a variance $V$. 
Because of the universality of the $M_{NSC}$-\mstar\ relation, and unless otherwise stated, we  perform the analysis on the combined dataset shown in  \myfig{fig:scaling}. 
We have verified that fitting only the NGVS data does not modify the results. 
We assume uniform priors for the parameters in closed intervals  as indicated in Table\,\ref{tab:fit}, and we keep $\gamma = 1$ because it is largely unconstrained by the data due to the small sample size at high masses \citep[see also][]{Grossauer2015}. 

\begin{deluxetable}{lrr}
\tablecaption{Bayesian fit to the \\ NSC-to-galaxy stellar mass relation \label{tab:fit}}
\tablehead{Parameter & Prior & Posterior}
\startdata
log\,($M^{'}_{NSC}/$\msun) 	& [5.0, 9.0] 	& $7.29\substack{+0.11 \\ -0.13}$ \\
log\,($M^{'}_{*}/$\msun) 	& [6.0, 11.0] 	& $9.73\substack{+0.10 \\ -0.12}$ \\
$\beta$ 			& [0.0, 10.0] 	& $0.46\substack{+0.03 \\ -0.04}$ \\
$\delta$ 			& [0.0, 10.0] 	& $0.43\substack{+0.23 \\ -0.17}$ \\
ln\,$V$ 			& [-5.0, 2.0] 	& $-1.83\substack{+0.09 \\ -0.09}$ 
\enddata
\tablenotetext{}{Note --The priors for the parameters listed above are assumed to be uniform within the closed intervals shown in the second column. The posterior figures correspond to the median of the marginalized posterior distribution and the associated 68\% confidence interval.}
\end{deluxetable}

The last column of Table\,\ref{tab:fit} shows the median and 68\% confidence interval for the marginalized posterior distributions of the parameters.
The corresponding best-fit relation is plotted as a solid line in \myfig{fig:scaling}, together with the 68 and 95 per cent confidence intervals. 
According to this fit, NSC masses scale  as as $M_{NSC} \propto$~\mstar$^{0.46}$ at the low mass end, a behavior that extends for over four decades in galaxy stellar mass.
This is a  shallower slope than what was found by \citet{denBrok2014} in Coma, $\beta = 0.57 \pm 0.05$, or by \citet{Scott2013} using a compilation from literature data, $\beta = 0.6 \pm 0.1$. 

The NSC-to-galaxy stellar mass relation features a characteristic mass for nuclei $M^{'}_{NSC} \approx 2\times10^{7}$ \msun\ in galaxies with stellar masses $M^{'}_{*} \approx 5\times10^{9}$ \msun. 
The ratio $(M^{'}_{NSC}/M^{'}_{*}) \approx 3.6 \times 10^{-3}$ that we infer is in perfect agreement with the mean NSC mass fraction obtained in the ACSVCS and ACSFCS studies \citep{Cote2006,Turner2012}, which can be simply explained  by their sample being dominated in number by intermediate-mass cluster galaxies.
We also measure an intrinsic scatter at fixed galaxy stellar mass of  $\sqrt{V} = 0.4$ dex, which is consistent with the value of $\approx$\,0.36 dex measured by \citet{denBrok2014} in Coma.
The value of $V$ is anticorrelated with the estimated uncertainties in the measurements of $M_{NSC}$, such that if the uncertainties are  underestimated then  $V$ would be smaller.
We interpret this large intrinsic scatter as indication that the process of NSC growth is quite stochastic (see Sect.\,\ref{sect:discussion}), and note that high stochasticity is a general characteristic of star formation processes in all low-mass halos \citep{Ricotti2016}.

\subsubsection{Relation to galaxy structure}
\label{sect:struct}
There are previous claims in the literature that nucleation and the properties of NSCs are related to galaxy structure.
For example, \citet{denBrok2014} found that at fixed galaxy luminosity, galaxies tend to have more luminous clusters when they have higher Sersic indices and rounder shapes.
The relation between galaxy structure and nucleation will be explored in detail in a different paper of this Series (C\^ot\'e et al., in preparation). 
Here we only point out that there is a tendency for nucleated galaxies to have slightly smaller effective radii at fixed galaxy luminosity, but the trend is not statistically significant considering the small sample size.
We will address this question in detail once the full NGVS sample is available.

\subsection{The NSC mass function in Virgo}
\begin{figure}[!t]
\includegraphics[angle=0,width=0.5\textwidth]{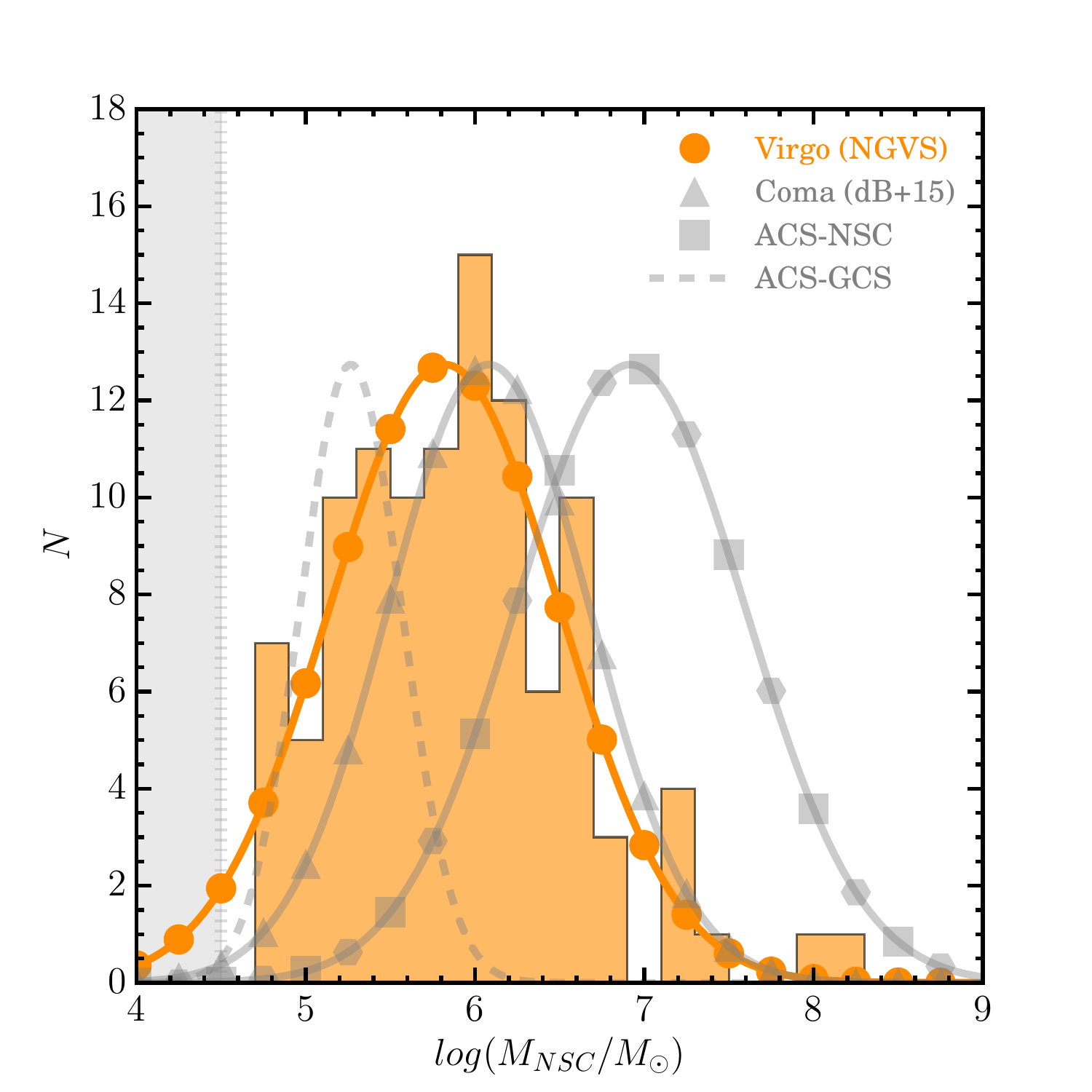}
\caption{The NSC mass function in the core of Virgo is shown here with a filled histogram. The best-fit Gaussian function is also plotted with circles, as well as the mass functions for other samples of NSCs and GCs from the literature. For reference, the typical NSC in Virgo is nearly four times more massive than the typical GC. The shaded vertical region indicates our 10$\sigma$ detection limit for NSCs.
}
\label{fig:massfunc}
\end{figure}

The histogram in \myfig{fig:massfunc} shows the NSC mass function (NSCMF) in the core of Virgo.
The solid line with circles shows the  best-fit Gaussian function with parameters $\mu = 5.82$\,dex and $\sigma = 0.68$. 
This functional form was selected for historical reasons and to facilitate a direct comparison with previous work on other compact stellar systems, such as GCs and UCDs.
As already noted by \citet{Turner2012}, the observed distribution is nothing but the convolution of the nucleation fraction with the nucleus-to-galaxy mass relation, and there is no obvious physical reason for this quantity to be normally distributed. 
The shaded vertical region in \myfig{fig:massfunc} indicates our 10$\sigma$ detection limit of log\,($M_{NSC}/$\msun)~$=4.5$ (see Sect.\,\ref{sect:scaling}). 
The sharp cutoff of the NSCMF at low masses suggests that we may be missing a small fraction of the nuclei below this mass limit, but the declining nucleation fraction toward low masses leads us to the conclusion that it has to be a very small number.

We compare the NSCMF in the NGVS with that derived from the \citet{denBrok2014} dataset for Coma early-type galaxies (triangle line). 
We transform their $i$-band magnitudes to stellar masses and obtain the best-fit Gaussian parameters $\mu = 6.08$ and $\sigma = 0.59$.
Thus the Coma NSCMF is very similar to ours, but it features a slightly more massive turnover mass which we attribute to the brighter cutoff in the Coma galaxy sample ($M_{i} \leq -14$ mag).
We also compare with the combined samples from the Virgo and Fornax ACS surveys \citep[][square-hexagon line]{Turner2012}. 
The bias is even more acute for these samples, which provide a very uniform sampling of the  massive end of the NSCMF, but lack galaxies fainter than $M_{g} \approx -15$, and therefore miss a large fraction of the faintest nuclei. 
Interestingly, a comparison with the GCMF \citep[][dashed Gaussian curve]{Jordan2007} shows that the average NSC is nearly four times more massive than the typical GC in Virgo galaxies. 
The two MFs naturally have very different widths, and it is perhaps not surprising that the faintest NSCs have masses virtually identical to those of the faintest GCs--which raises the question, how different are these NSCs from ordinary faint GCs?


\section{Discussion}
\label{sect:discussion}

The present study contains three important results pertaining to the formation of NSCs and the relation to their host galaxies. 
First, to a high degree it is galaxy mass that regulates NSC formation and growth, in the sense that both very massive and very faint galaxies have low likelihoods of hosting nuclei.
However, a secondary dependence on environment seems to indicate that the efficiency of NSC formation is also affected to a some degree by the mass of the host halo, such that at fixed stellar mass galaxies in denser environments have a higher probability of being nucleated. 
Finally, even if environment somewhat regulates the likelihood of a given galaxy to host an NSC, the universality of the NSC-to-galaxy mass relation indicates that the mass of the central star cluster is primarily controlled by the galaxy mass--albeit with a large scatter, which probably is an indication of the stochastic nature of NSC growth.
We now turn to discuss the implications of these results in the context of NSC formation scenarios.

\subsection{The connection between NSCs, GCs, and DM halo mass}

\begin{figure}[!t]
\includegraphics[angle=0,width=0.5\textwidth]{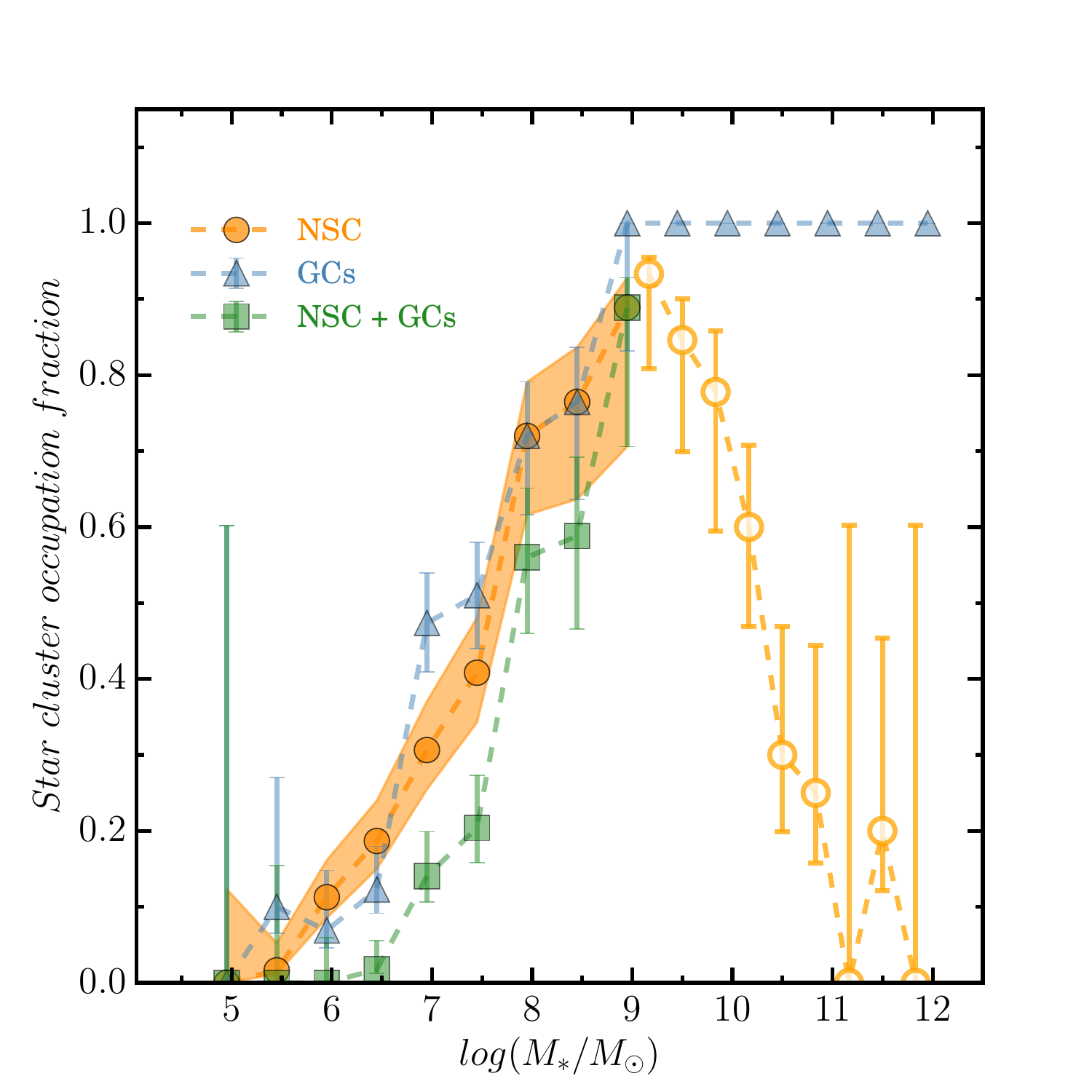}
\caption{Star cluster occupation distribution in the core of  Virgo  as a function of galaxy stellar mass. The fraction of galaxies hosting an NSC, GCs, and the two types of clusters simultaneously are represented with circles, triangles, and squares, respectively. Shaded regions and error bars indicate the corresponding 68 per cent Bayesian credible interval. The remarkable similarity between these occupation distributions indicates a close connection between the different families of compact stellar systems.}
\label{fig:frac_gcs}
\end{figure}

As has been previously discussed, it is natural to wonder about the relationship between the different families of star clusters in galaxies, and what is the role of total galaxy mass in the establishment of these relations. 
This is particularly true for GCs and NSCs, because the early decay and merging of dense star clusters seems a viable--perhaps even unavoidable--mechanism in the centers of galaxies. 
Also, the total mass of the GC system (GCS) has been shown to correlate with the dark matter (DM) halo mass of the galaxy \citep{Peng2008,Spitler2009,Georgiev2010,Hudson2014}, and one may naively expect a similar behavior for other compact stellar systems.

\subsubsection{NSCs and GCs}
In \myfig{fig:frac_gcs} we reproduce the NSC occupation distribution in Virgo, but now also include the fraction of galaxies that host GCs (triangles). 
The GC candidates are selected from the mixture model classification by Peng et al. (in prep.), and their numbers counted to $g < 24.5$ ($\sim$\,0.6 mag fainter than the GCLF turnover magnitude in Virgo) within a distance $<$\,2.5\,$R_{e}$ from the center of the galaxy (excluding the NSC). 
A local background level was determined in an annulus between 4 and 10\,$R_{e}$, and subtracted to account for the contamination by stellar and extragalactic interlopers, and by the rich GCS of M87.
The final number $N_{GC}$ of globular clusters is corrected to the full GCLF.
Consistent with previous work, \myfig{fig:frac_gcs} shows that all galaxies more massive than $M_{\star} \approx 10^{9}$ \msun\ host at least one GC \citep{Peng2008, Georgiev2010}.
Remarkably, the fraction of galaxies containing GCs decreases toward lower masses at exactly the same rate as the nucleation fraction. 
Within the uncertainties, the two curves are indistinguishable from each other, which lends support to the idea that these two types of stellar systems are closely interconnected--perhaps both being the result of similar underlying physical processes. 

While  the GC and NSC occupation distributions are identical from a statistical point of view, this is not entirely the case on a galaxy per galaxy basis. 
In the same figure we show with squares the fraction of galaxies that simultaneously host an NSC {\it and} GCs. 
As before, here we do this exercise for galaxies with stellar masses ranging from $M_{*} = 10^{5}$ \msun\ to $M_{*} = 10^{9}$ \msun, and all figures discussed in this section are only valid for objects within this mass range. 
The NSC+GC occupation distribution  is slightly different than the NSC one, in the sense that at fixed galaxy stellar mass there is a lower fraction of galaxies that host the two types of star clusters. 
The breakdown of the star cluster occupation fraction in the core of Virgo is as follows: 43\% of the galaxies in the mass range under study do not host any type of compact star cluster; 18\% are non-nucleated galaxies that have GCs; 22\% have both an NSC and GCs (the squares in \myfig{fig:frac_gcs}); and 17\% of the galaxies have NSCs but no GCs.
Of course, with the current dataset we can not rule out the possibility that the latter subpopulation at some point {\it did have} GCs that  have been stripped during the orbital evolution within the cluster, or that their GCs merged to form the NSC.
A perhaps even more intriguing possibility is  that the NSC is indistinguishable from a regular old GC located at the bottom of the potential well. 
Detailed studies of the stellar populations in these low-mass NSCs are required to explore this hypothesis. 

\begin{figure}[!t]
\includegraphics[angle=0,width=0.5\textwidth]{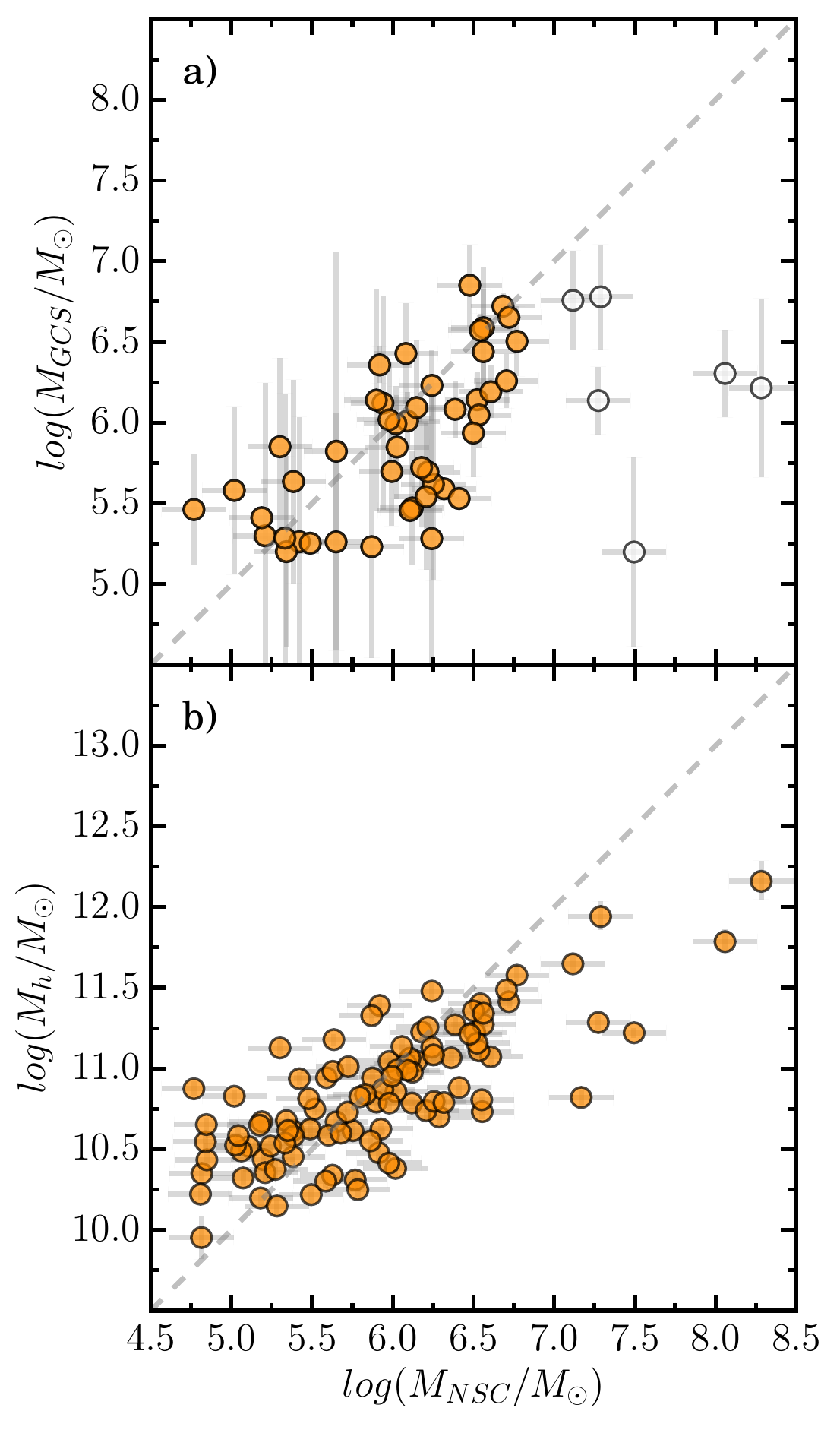}
\caption{The masses of NSCs in the core of the Virgo cluster. 
{\it Top}: The estimated total mass of the GCS  within 2.5\,$R_{e}$ is shown against the mass of the nucleus. 
The two quantities are consistent with each other for masses $M_{NSC} < 10^{7}$ \msun.
The number of GC candidates is artificially biased low in more massive galaxies because of the aperture used to estimate their numbers (open symbols; see text for details). 
{\it Bottom}: The ordinates show the peak DM halo mass for each galaxy from abundance matching plotted versus the mass of the nucleus. 
The two masses are correlated, but the relation differs from the constant mass fraction (shown as a dashed line for $M_{h}/M_{NSC} = 10^{5}$).}
\label{fig:all_masses}
\end{figure}

In \myfig{fig:all_masses}a we also study what is the relation between the mass of the NSC and the total mass of the GCS ($M_{GCS}$), which we derive using the following Monte Carlo method. 
For each nucleated galaxy we randomly sample a normal distribution with mean value $N_{GC}$ and standard deviation $\Delta N_{GC}$, where these quantities are extracted from the GC candidate catalog.
The resulting number of GCs is rounded to the closest integer number $Z_{GC}$, and if the galaxy has at least one GC ($Z_{GC} \ge 1$) we assign them masses according to a normally-distributed GCMF.
\citet{Jordan2007} show that the GCMF is not universal, but rather exhibits a dependence on the luminosity/stellar mass of the host galaxy such that fainter systems feature lower turnover masses and narrower MFs. 
Therefore, for each nucleated galaxy in the NGVS we draw GCs from a normal GCMF with turnover masses and logarithmic dispersions consistent with the (extrapolated) relation shown in Fig.\,14 from \citet{Jordan2007}.
The masses of the $Z_{GC}$ clusters are summed to obtain a total mass for the GCS, and this Monte Carlo process is repeated 10,000 times for each galaxy. 
We record the mean GCS mass and its standard deviation, which are plotted against the masses of NSCs in the top panel of \myfig{fig:all_masses}.
We note that at low $M_{GCS}$ values there is an unavoidable level of discreteness in the mean masses caused by galaxies that are consistent with having a single GC. 
In these cases the mean mass simply corresponds to the turnover mass of the GCMF evaluated at the corresponding galaxy stellar mass.
The properties of the GCSs in the core of Virgo from NGVS data will be discussed in more detail in future papers of this series. 

\myfig{fig:all_masses}a shows that $M_{NSC}$  and $M_{GCS}$ track each other remarkably well for masses $M_{NSC} < 10^{7}$ \msun, which corresponds to  $M_{*} < 3\times10^{9}$ \msun. 
In comparison more massive galaxies seem to feature depleted GCSs, but this is an artefact caused by the way the number of GCs is estimated. 
As described above, GC candidates are only selected within a projected distance $R < 2.5 R_{e}$ from the center of the galaxy. 
While this is distant enough to account for the majority of GCs in low-mass systems, more massive galaxies certainly have GCSs extending far beyond this limit. 
As a result, we are progressively missing a larger fraction of the GCS as we move toward larger masses, and hence the apparent bend in the trend displayed in \myfig{fig:all_masses}a.

In summary, for low mass galaxies the data once again show a close connection between the two types of compact stellar systems, now in terms of their total masses. 
We conclude that while the presence of GCs is not a sufficient condition to form an NSC in low-mass galaxies \citep{Miller1998,denBrok2014}, {\it the  two families of  star clusters probably simply are different manifestations of the prevalent mode of  star formation at early times}.
Indeed, both the star cluster occupation fractions and the relation between the total masses of GCSs and NSCs can be qualitatively understood under a scenario where only the galaxies that happen to form enough proto-GCs within the dynamical friction cone develop an NSC--which, in turn, grows proportionally to the size of the GC population. 
We will develop further this idea in Section\,\ref{sect:bias}.

\subsubsection{NSCs and DM halo mass}
We now attempt a comparison between  $M_{NSC}$ and  the DM halo masses, $M_{h}$.
To determine $M_{h}$ we make use of the SHMR obtained by \citet{Grossauer2015} for the NGVS via abundance matching. 
This technique relies on the assumption that there exists a univocal relation between stellar and DM halo masses, such that the $n$th-ranked galaxy occupies the $n$th most massive halo.
We warn that this approximation most likely breaks down at the lowest masses probed by the NGVS due to the inefficiency of galaxy formation at these scales. 
For reference, \citet{Fattahi2016} show that in the APOSTLE simulation half of the halos with masses $M_{h} = 10^{9.5}$ \msun\ remain fully dark. 
This in turn results in a systematic offset between the actual halo masses of the simulated galaxies and the $M_{h}$ that would be inferred from the different abundance matching models. 
We finally note that because the Virgo sample is essentially comprised of satellites, the matching procedure is carried out at peak halo mass, and accordingly $M_{h}$ represents the maximum DM halo mass ever attained by the galaxy--which occurs at the redshift of infall and not at $z=0$. 
Present-day $M_{h}/M_{*}$ ratios are expected to be much smaller due to preferential stripping of DM halos relative to the stellar component as galaxies orbit within the cluster potential well \citep[e.g.,][]{Smith2013a,Smith2015}. 

With all these caveats in mind, in the lower panel of \myfig{fig:all_masses} we show the NSC masses in the core of Virgo plotted against the estimated $M_{h}$. 
Uncertainties in the stellar mass determinations and  in the SHMR are plotted as error bars in \myfig{fig:all_masses}.
This panel shows that there is a linear relation between log\,$M_{NSC}$ and log\,$M_{h}$, but it deviates from the constant mass fraction relation.
For reference, the diagonal dashed line shows the expected relation if nuclei and halo masses were offset by a constant mass ratio $M_{h}/M_{NSC} = 10^{5}$.
This difference actually is a direct result of the different low-mass slopes for the SHMR and the NSC-to-galaxy stellar mass relation. 
The abundance matching exercise by \citet{Grossauer2015} indicates that $M_{h} \propto M_{*}^{0.39}$, whereas we obtain $M_{NSC} \propto M_{*}^{0.46}$. 
Hence, we find a stronger dependence of the mass of the nucleus on halo mass,  $M_{NSC} \propto M_{h}^{1.2}$.
We note that $\beta$, the SHMR slope at the low mass end, is still poorly constrained and its value is highly debated in the literature.
Nevertheless, most recent works find it to be in the range $0.3 \lesssim \beta \lesssim 0.45$ \citep[e.g.,][]{Behroozi2013,Moster2013,Brook2014,Sawala2015}, and therefore the claim that NSC masses depend strongly on DM halo mass seems robust--provided the abundance matching approximation remains valid.

\subsection{Comparison with models for NSC formation}

We now interpret these results in the light of current models for NSC formation, and explore to what extent  they reproduce the observations in Virgo. 
Specifically, we compare the NGVS data against the  models from \citet[][hereafter M06]{McLaughlin2006} and  \citet[][hereafter A15]{Antonini2015}.
M06 present a fully analytic model for the self-regulated growth of NSCs, where  feedback from stellar winds and supernovae drive a superwind from the nucleus with a momentum flux directly proportional to the Eddington luminosity.
When the NSC reaches a critical mass the superwind can escape the galaxy, thus halting accretion and freezing $M_{NSC}$.
A15, on the other hand, make predictions for both the nucleation fraction and the NSC-to-galaxy stellar mass relations within a cosmological framework for galaxy formation.
Their model comes in two flavours.
The first one simply consists of a purely dissipationless process, whereby star clusters migrate to the galactic center and merge under the effect of dynamical friction (hereafter the {\tt CliN} model). 
It is therefore very similar in spirit to \citep{Gnedin2014}, but like most of these models it suffers from the limitation that it does not capture the hierarchical buildup of galaxies nor any dissipative process related to star formation.
The second model ({\tt GxeV}) addresses the latter aspects, as it first follows the growth and merger histories of galaxies and their DM halos, and then incorporates a recipe for the formation of NSCs with a treatment for dissipative processes leading to nuclear star formation. 
This proved to be an important element, because according to their calculations more than half of the total NSC mass is accounted for by in situ star formation, and dissipative processes appear to become increasingly important with galaxy mass.
The A15 models also take into account the disruptive effects of MBHs, which will become relevant when addressing NSC occurrence in high-mass galaxies.

\subsubsection{The NSC occupation distribution}

\begin{figure}[!t]
\includegraphics[angle=0,width=0.5\textwidth]{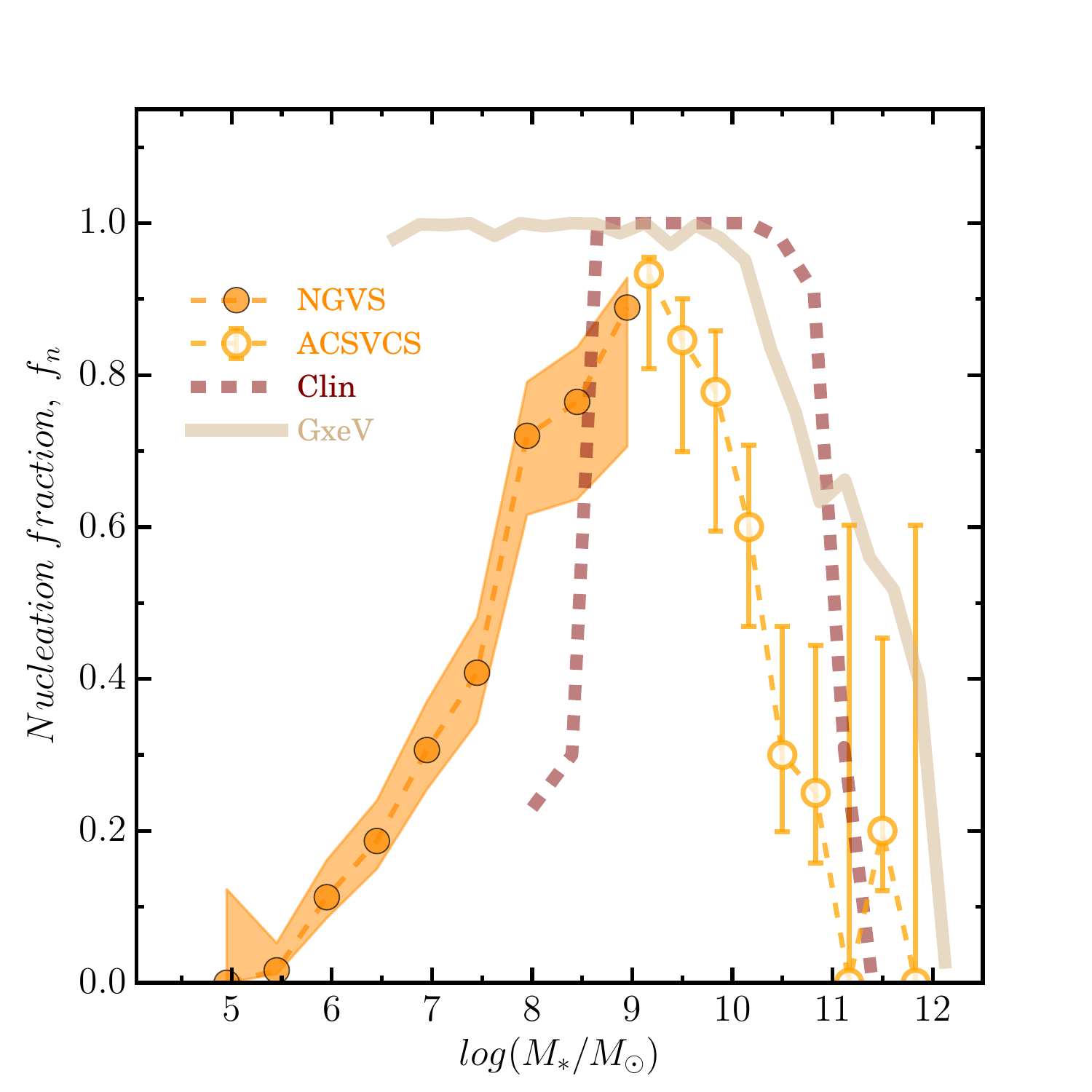}
\caption{Same as \myfig{fig:frac_gcs}, but now the lines show the predictions for the NSC occupation fraction from the models by \citet{Antonini2015}. The dashed line corresponds to a purely dissipationless model ({\tt CliN}) of NSC growth via dynamical friction-driven mergers of pre-existing dense star clusters. The solid line is for a dissipative semianalytic model ({\tt GxeV}) that incorporates gas inflows and in situ star formation. See text for discussion.}
\label{fig:frac_sims}
\end{figure}

In \myfig{fig:frac_sims} we again reproduce the nucleation fraction in Virgo, now in direct comparison with the two sets of models from A15 (M06 do not make predictions for this quantity). 
The dissipationless {\tt CliN} scenario (dashed line) is able to reproduce, at least qualitatively, the peaked form of the NSC occupation distribution. 
However, it predicts too high of an efficiency for NSC formation at intermediate-to-high galaxy masses.
At the same time it underpredicts the nucleation fraction for masses below log(\mstar/\msun$)\approx8.5$.  
The  solid line corresponds to the NSC occupation fraction in the dissipative model. 
Here in situ star formation contributes significantly to the growth of NSCs in massive galaxies, and therefore the nucleation fraction departs even more from the observed behavior. 
If taken at face value the results at the low mass end are even more discouraging for the {\tt GxeV} model, which predicts $f_{n} \approx 100\%$ for all masses $M_{\star} \lesssim 10^{10}$ \msun.
Yet the observations indicate a steady decline of $f_{n}$ toward low masses.
However, A15 warn against putting too much trust in these numbers, because the identification of NSCs against the galaxy background in the dissipative {\tt GxeV} model is poor and the algorithm does not follow the evolution of individual star clusters, but only the average nuclear mass infall rate.

The sudden drop in $f_{n}$ at high masses \mstar $\gtrsim 10^{9}$ \msun\ is a well-known result that is traditionally interpreted in the literature as the result of cluster disruption by the MBHs that inhabit the galactic centers. 
We note that the onset of this decline is perfectly consistent with recent results indicating that the MBH occupation fraction is high above \mstar $\sim 10^{9}$ \msun\ \citep{Miller2015,Nguyen2017}.
In the A15 models MBHs contribute to halt NSC formation and growth in two important ways. 
First, their strong tidal field enhances mass loss from star clusters as they decay toward the center, while simultaneous dynamical heating decreases their binding energy. 
This all contributes to a significant reduction in the amount of mass deposited in the NSC, if not implying the full disruption of the inspiraling star clusters. 
It is not clear at this stage why the Virgo galaxies seem to have a lower NSC occupation fraction at high masses compared to the models.
One possibility is that the semianalytic formalism fails to capture the structural nonhomology of early-type galaxies. 
\citet{Emsellem2007} show that  for density profiles with S\'ersic indices $n \gtrsim 3.5$ tidal forces become disruptive nearly everywhere, and therefore hinder the efficient collapse of gas and its subsequent transformation into stars. 
Given the observed $n$-\mstar\ relation in Virgo \citep[][]{Ferrarese2006}, this mechanism should operate in a majority of the early-type galaxies more massive than \mstar~$\approx10^{10}$~\msun.
Alternatively, it is possible that the difference is an effect associated to the efficiency of star cluster disruption by MBHs as implemented in the semianalytic model.

For low mass galaxies, however, this framework is unlikely to apply for two reasons. 
First, in several scenarios of MBH formation the occupation fraction is expected to be a relatively strong increasing function of galaxy mass \citep[][and references therein]{Volonteri2010}. 
Many of these low mass galaxies probably do not host a MBH at all. 
But even if they do its mass would need to be larger than $M_{\bullet} \approx 10^{8}$ \msun\ to efficiently disrupt inspiraling star clusters and halt any further NSC growth (see the discussion in A15 about the importance of this characteristic mass scale). 
According to the observed $M_{\bullet}$-\mstar\ relation \citep[][]{McConnell2013}, this value is well above the expected BH masses in \mstar~$\lesssim 10^{9}$~\msun\ galaxies.

Galaxies in the core of Virgo with stellar masses below this limit have light profiles well described by S\'ersic indices in the range $0.5 < n \lesssim 1.5$ (Ferrarese et al. 2016b) which, as \citet{Emsellem2007} show, feature compressive tidal forces in the central regions that are conducive to NSC formation.
And yet the observations indicate that fewer lower mass galaxies eventually form nuclei. 
Possible solutions to this puzzle range from stellar feedback preventing cold gas from reaching the nuclear regions of these galaxies \citep[][]{El-Badry2015},  to the presence of very cuspy halos such that DM is a dominant mass component in the very central regions of these galaxies.

We propose instead that a more likely explanation for the paucity of NSCs toward low galaxy masses simply is  a low initial number of dense star clusters.
We have shown that the existence of GCs is tightly linked to the presence of NSCs, and numerous studies have now established that the total mass of the GCS correlates tightly with the DM mass of the galaxy \citep{Peng2008,Spitler2009,Georgiev2010,Hudson2014}. 
If this holds for very low mass galaxies and they form a low number (but high mass fraction) of bound star clusters then it is natural to expect that many faint galaxies simply lack the ingredients to form an NSC seed in the first place.

\subsubsection{The NSC-to-galaxy stellar mass ratio}

\begin{figure}[!t]
\includegraphics[angle=0,width=0.5\textwidth]{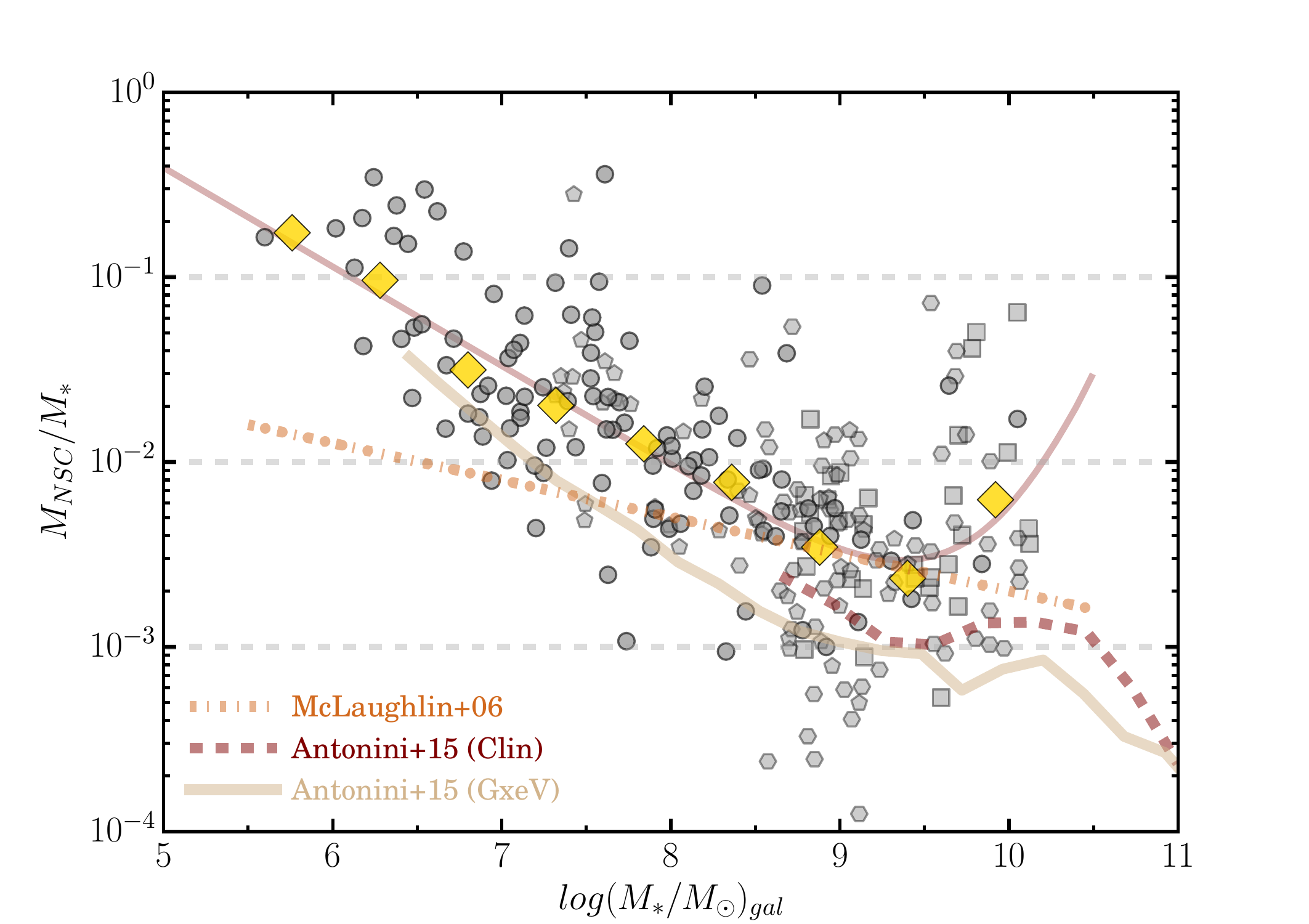}
\caption{The NSC-to-galaxy stellar mass ratio for the NGVS and literature samples. Symbols are as in \myfig{fig:scaling}. The large diamonds show the mean relation in bins of stellar mass, and they highlight strong dependence of the nuclear mas fraction on host mass. The thin solid line corresponds to the best-fit $M_{NSC}$-\mstar\ relation. The mass ratio spans nearly three orders of magnitude, with the most prominent nuclei being almost 50\% as massive as their host galaxies. The thick lines correspond to different predictions from a dissipative (solid) and two different dissipationless models for NSC formation (dashed and dot-dashed lines). See text for details.}
\label{fig:nucmass_sims}
\end{figure}

In \myfig{fig:nucmass_sims} we show the NSC-to-galaxy mass ratio as a function of galaxy stellar mass for the NGVS and the literature samples, with symbols as indicated in the legend of \myfig{fig:scaling}.
These two  figures are essentially equivalent, but \myfig{fig:nucmass_sims} does a better job at highlighting the increasing prominence of NSCs toward lower masses. 
As in \myfig{fig:scaling}, horizontal dashed lines indicate constant NSC mass fractions of 0.1\%, 1\% and 10\%, from bottom to top respectively.
To better illustrate the dramatic dependence of the nuclear mass fraction on host stellar mass, large diamonds show the mean $M_{NSC}/$\mstar\ ratio in bins of constant galaxy stellar mass.
The mass ratio spans nearly three orders of magnitude, with a mean value that drops to $\approx$\,0.36\% for galaxies with \mstar\ $\approx 3\times10^{9}$ \msun\, and then increases  for both more and less massive galaxies.
This minimum value for  $M_{NSC}/$\mstar\ is very similar to the constant mass ratio found in the Virgo  and Fornax ACS surveys \citep{Cote2006, Turner2012}.
But \myfig{fig:nucmass_sims} indicates that the ratio is anything but constant. 
The increasing prominence of NSCs is exacerbated at the lowest mass scales, where a few nuclei are a whopping $\approx$\,50\% as massive as their hosts. 
The trend is present in all the samples of cluster low-mass galaxies included in \myfig{fig:scaling}, but it is only thanks to the significant extension toward the low-mass end enabled by the NGVS that we can assess its statistical significance with confidence. 
We will now discuss these results in the context of dissipative and dissipationless models of NSC formation. 
\new{More specifically, there is a clear prediction from models of NSC growth via GC inspiraling that the nuclear mass fraction should scale as $M_{NSC}/M_{*}\propto M_{*}^{-0.5}$ \citep{Antonini2013,Gnedin2014}. 
A simple inspection of  \myfig{fig:nucmass_sims} indicates that this behavior does not hold at high galaxy masses, and this is a clear indication that additional (dissipative) mechanisms have to be invoked to explain these NSC masses.} 

The thick dot-dashed line in \myfig{fig:nucmass_sims} shows the prediction for the $M_{NSC}/$\mstar\ ratio from the feedback-regulated model by M06.
We use their Eq. 10 with parameters $\lambda = 0.05$ for the wind thrust efficiency, $Z = 0.5\,Z_{\odot}$ for the stellar metallicities and $v_{w} = 200$ km\,s$^{-1}$ for the velocity of the superwind.  
These values were chosen to reproduce the NSC mass fraction at the intermediate-mass regime where the kink of the  $M_{NSC}/$\mstar\ ratio occurs, but it is clear that at lower masses the analytical relation is too shallow compared to the data.
This is because in this model the self-regulated mass of the nucleus scales as  $M_{NSC} \propto M_{*}^{-1/5}$, which is inconsistent 
with our best-fit slope (cf. Table\,\ref{tab:fit}).
It follows that if feedback played a relevant role in setting the initial masses of NSCs, additional growth mechanisms are required to explain their present-day stellar content.

Solid and dashed thick lines in \myfig{fig:nucmass_sims} correspond to the predictions for the $M_{NSC}/$\mstar\ ratio from the dissipative  ({\tt GxeV}) and the dissipationless ({\tt CliN}) models by A15, respectively.
It is interesting--or perhaps worrying--that  these two models seem to reproduce the observed trend at low masses.
While predicting the correct slope for the $M_{NSC}$-\mstar\ relation can be seen as a significant success, it also means that  this  relation provides little to no  discriminating power on possible formation scenarios. 
The similarity between the dissipationless and the dissipative models indicates that, essentially, {\it at low galaxy masses NSC growth is controlled by the average nuclear mass infall rate, independent of whether it is constituted by stellar or gaseous material.}
The stellar population properties of the NSCs {\it do} differ in the dissipative and the dissipationless scenarios, and detailed studies of the least massive NSCs will clarify if the complex star formation histories found in the nuclei of intermediate-mass galaxies \citep{Monaco2009,Seth2010,Paudel2011} are mirrored at the smallest scales.

As already mentioned, the very low mass ratio predicted by all these models for \mstar\ $\gtrsim 10^{10}$ \msun\ galaxies is troubling.
However, these massive NSCs in Virgo and Fornax are quite peculiar systems that do not resemble the nuclei found in lower mass objects in a number of ways. 
\new{They feature more flattened morphologies, very large half-light radii and their colors show increased scatter compared to lower mass systems--with a predominance of nuclei that are even redder than their host galaxies \citep{Cote2006,Turner2012}.
In light of these properties it has been proposed that these massive NSCs resemble the 'dense stellar cores' that form in some numerical simulations as a result of dissipative processes involving mergers and/or nuclear gaseous inflows \citep{Mihos1994a}.}
The observed discrepancy between data and models  implies that the formation mechanisms of such a mass excess in the central regions of the most massive galaxies are not fully captured by current NSC formation models, even when they incorporate dissipative processes.

Overall, the shape of the NSC-to-galaxy stellar mass ratio indicates the existence of two well-defined mass regimes. 
Below \mstar\ $\approx 5\times10^{9}$  \msun\ NSCs become increasingly prominent, and we will now discuss this finding in the context of dynamical friction-driven coalescence of dense star clusters.

\subsection{A scenario for biased NSC formation}\label{sect:bias}
We have shown that the present-day  galaxy mass (stellar or total) seems to be the main parameter controlling not only whether a galaxy harbors an NSC, but also its subsequent growth.
This is in line with previous work, but our results further  demonstrate that this simple picture is incomplete. 
Our finding that at fixed stellar mass the nucleation fraction shows a secondary dependence on the mass of the host halo indicates that NSC occurrence is a more complex phenomenon that depends on properties related to the global environment. 
We are not aware of any model for NSC formation that reproduces this effect in early-type galaxies, but the result is very reminiscent of the discovery by \citet{Peng2008} that the GC specific frequencies in Virgo galaxies are also biased toward dense environments. 
Specifically, the average GC mass fraction for \mstar\ $\lesssim 5\times10^{9}$ \msun\ galaxies increases from the galaxy cluster outskirts to the center.  
\citet{Peng2008} were able to show that, at least qualitatively, the trend can be explained by the preferential formation of GCs in the earliest collapsing halos that can efficiently form stars before reionization.
In this biased scenario, the galaxies that inhabit the central cluster regions would have collapsed first, starting to form stars earlier and did so at higher star formation rates (SFRs) and higher star formation surface densities ($\Sigma_{SFR}$) than the systems that are presently found in the cluster outskirts.
Old GCs are believed to form precisely in regions featuring high $\Sigma_{SFR}$ and enormous pressures \citep{Harris1994,Elmegreen1997,McLaughlin1999b,Ashman2001,Kruijssen2015,Pfeffer2018}.
This, together with the fact that satellites residing in higher density environments are accreted at earlier times and formed stars rapidly \citep{Liu2016}, naturally results in higher present-day GC mass fractions for the innermost systems. 

Here we speculate that the same biased formation channel for star clusters  is responsible for the observed environmental dependence of the nucleation fraction toward low galaxy masses.
It is well established that the clustering of DM halos is a strong function of mass and formation time, especially for ancient, low-mass halos \citep{Gao2005,Wechsler2006,Dalal2008,Lacerna2011}. 
This is equivalent to saying that at fixed peak mass subhalos form earlier in more massive host halos, an effect known as assembly bias.
All the low-mass cluster galaxies shown in \myfig{fig:frac} inhabit regions with similarly high mean overdensity ($R/R_{vir} \lesssim 0.25$), and are therefore expected to be the most ancient population in these clusters.
For reference, subhalos in virial equilibrium in the cores of massive clusters  have typical infall times of $t_{inf} \gtrsim$\,6 Gyr \citep{Oman2013}.
Thus at fixed galaxy stellar mass all these subhalos did reach a similar peak halo mass,\,\footnote{This statement is of course only valid for galaxies that have not suffered significant stellar mass loss after infall. While this is generically true for the average satellite population, \citep{Watson2013}, it may be less valid for these low-mass galaxies in the cores of massive clusters \citep{Smith2015}.} but those in the core of Coma attained it at earlier times than those in the cores of Virgo (or Fornax). 
In fact, prior to infall the subhalos in denser environments were {\it at all times} more massive than those in slightly less dense regions.

If we extrapolate back in time to the epoch of GC formation ($z \gtrsim 2$) we find a scenario equivalent to that proposed by \citet{Peng2008}, namely that low-mass galaxies in Coma started forming stars earlier than in Virgo, and at higher SFRs and $\Sigma_{SFR}$. 
These conditions were conducive to the formation of bound young massive clusters (YMCs), and if cluster formation efficiency was close to universal \citep{McLaughlin1999b} and galaxies formed YMCs proportionally to their mass at that epoch \citep{Kruijssen2015}, then one naturally expects a larger mass fraction in star clusters in the more biased (proto-)Coma galaxies.\,\footnote{Note that if this scenario is correct, it also implies that the low-mass galaxies in clusters like Coma must feature higher GC mass fractions (or specific frequencies) than those in Virgo.} 

The YMCs that were born closer to the center of the potential well and survived the early disruption phase \citep{Kruijssen2011}  were able to merge within a few dynamical times.
Simple dynamical friction arguments  \citep[e.g.,][]{Lotz2001} indicate that, for  \mstar\ $\lesssim 10^{8}$ \msun\ galaxies, a star cluster with the GCMF turnover mass would decay from a distance of $\sim$\,1 kpc in less than a few billion years.
In this context it is important to recall that DM halos  grow their central potential very rapidly at early times, but during the subsequent long-lasting slow accretion regime the material builds up predominantly in the outskirts and the central densities change very little \citep{Wechsler2002,vandenbosch2014}.
Under these conditions, and considering the very low masses of the faintest nucleated galaxies in Virgo--almost comparable to those of regular GCs--the dynamical friction-driven orbital decay of bound star cluster probably has been a very efficient mechanism throughout a large fraction of the galaxy history.

Baryons of course complicate this simple picture, and dissipative processes have probably contributed to some extent to the growth of NSCs in low-mass galaxies.
The mass distribution of (non-cuspy) low-mass galaxies favors nuclear gas inflows \citep{Emsellem2007}, and with all likelihood the last star formation events in these quiescent galaxies took place in the central regions.

Admittedly, this scenario is highly speculative. 
But the proposal that nucleated early-type cluster galaxies are a biased subpopulation is actually not new, but rather consistent with previous results in the literature.
These galaxies have a tendency to inhabit the inner and higher density cluster regions \citep{Binggeli1991,Lisker2007}; their velocity distribution reveals a preference for circularized orbits compared to  \citep{Lisker2009}; and there is tentative evidence that they feature higher GC mass fractions than non-nucleated cluster galaxies \citep{Miller1998,Sanchez-Janssen2012}.
\new{The results presented here just add another element in support of this picture, but many questions remain unanswered. 
For example, why does this scenario result in $\sim50\%$ of all the mass in old star clusters ending up in the NSC regardless of galaxy mass? What is the nature of  the observed change in the mass fraction slope at high masses? If a larger fraction of stars in cluster galaxies indeed form in bound star clusters, the na\"ive expectation is that the NSC mass fraction should depend on environment, but this is not observed. Finally, this scenario must also provide an explanation for the mildly bluer colors and younger ages of NSCs relative to their host galaxies \citep[e.g.][]{Spengler2017}.
}

\subsection{UCDs, satellite disruption, and mass deposition on stellar halos}

\begin{deluxetable}{lrrr}
\tablecaption{Inferred stellar masses for the progenitors of stripped NSC candidates in the Local Group\label{tab:galmass}}
\tablehead{Cluster & $M_{GC}$/\msun\ & \mstar$_{,p}$/\msun\ & Ref.}
\startdata
$\omega$Cen & 2.5$\times10^{6}$ & 6.3$\times10^{8}$ & (1) \\
M54 & 2.0$\times10^{6}$ & 4.0$\times10^{8}$ & (1) \\
NGC\,2419 & 1.3$\times10^{6}$ & 1.9$\times10^{8}$ & (1) \\
M19 &  0.9$\times10^{6}$ & 0.9$\times10^{8}$ & (1) \\
\hline
Mayall\,II/G1 & 4.6$\times10^{6}$ & 15.0$\times10^{8}$ & (2) \\
G78 & 2.4$\times10^{6}$  & 5.9$\times10^{8}$ & (3) \\
G213 & 1.8$\times10^{6}$  & 3.6$\times10^{8}$ & (3) \\
G280 & 2.2$\times10^{6}$  & 5.0$\times10^{8}$ & (3) 
\enddata
\tablenotetext{}{ {\footnotesize(1) \citet{Mackey2005}; (2) \citet{Meylan2001}; (3) \citet{Fuentes-Carrera2008}.}
}
\end{deluxetable}

Soon after their discovery UCDs were already suspected to be the surviving nuclei of disrupted satellites \citep{Drinkwater2003,Bekki2003,Goerdt2008}.
Subsequent work has shown that they probably constitute a mixed bag of objects with galactic and star cluster origins, and that there is a trend for NSCs (GCs) to be dominant at the high (low) mass end \citep{Hasegan2005,Evstigneeva2007,Mieske2008,Mieske2012a,Norris2014,Zhang2015,Pfeffer2016}. 
An important link between UCDs and nucleated galaxies was recently discovered by \citet{Liu2015b} using NGVS imaging.
They present evidence that a fraction of the UCDs surrounding M87 and M49 are embedded in low surface brightness envelopes whose prominence correlates with the distance to these massive galaxies.
Thus the innermost UCDs show no evidence for such stellar halos, whereas the envelopes surrounding the most distant systems are so prominent that they unambiguously are nucleated galaxies. 
The nature of the intermediate objects in this morphological sequence is unclear at the moment, but the progression toward less prominent envelopes with galactocentric distance is consistent with the tidal   stripping picture \citep[e.g.,][]{Pfeffer2013}.
Alternatively, the objects with envelopes may simply be nucleated galaxies with very prominent NSCs like those we find at the low mass end.
However, we note that (i) there is no overlap between the samples of UCDs from \citet{Liu2015b} and our nucleated low-mass objects; and (ii) unlike with the UCDs, we do not find a correlation between the distance to M87 and the prominence of the NSC.

We note in passing that the recent hydrodynamic cosmological simulations by \citet{Ricotti2016} offer an intriguing alternative for the origin of these envelopes.
These authors follow the formation of very low-mass galaxies (\mstar\ $\lesssim 10^{6}$ \msun) before reionization, and in these simulations the majority of the stars form in dense star clusters.
The more massive clusters remain bound after gas is expelled by ongoing star formation, but those with masses $ \lesssim$\,$10^{4}$ \msun\  dissolve and expand until they become bound by the DM halo.
If the surviving more massive clusters merge driven by dynamical friction, the object left behind would closely resemble a (low-mass) UCD with an envelope \citep[see also][]{Milosavljevi2014}.

Coming back to the threshing scenario, \citet{Ferrarese2016} show that the observed $M_{NSC}$-\mstar\ relation can be inverted and used  in combination with the abundance of UCDs to provide an estimate of the amount of intracluster light in the core of Virgo contributed by disrupted satellites.
An even simpler and direct application of this relation is to compute the mass in stars contributed to the stellar halos of other massive galaxies by the putative progenitors of UCDs. 
The three most massive UCDs in the Virgo cluster are not located close to M87, but reside in the infalling group dominated by the massive early-types M60 and M59 \citep{Chilingarian2008a,Sandoval2015,Liu2015a}.
The  stellar masses of these UCDs are in the 0.5-1.5$\times10^{8}$ \msun\ range, and if they are the remnants of threshed progenitors that followed the observed mean relation in \myfig{fig:scaling} then their parent galaxies had stellar masses in excess of \mstar\ $\sim 10^{10}$ \msun.
M59 hosts the first and third most massive of the UCDs, and under this scenario  approximately 15-25\% of its stellar mass would have been accreted in two distinct merger events. 
This figure is consistent with recent cosmological hydrodynamic simulations of stellar halo formation, which  indicate a larger contribution  of accreted mass in the stellar halos of more massive galaxies \citep{Rodriguez-Gomez2016}.

Given the observed universality of the $M_{NSC}$-\mstar\ relation a similar exercise can be attempted in the Local Group, where some of the most massive GCs have long been suspected to be a distinct subpopulation.
Specifically, $\omega$Cen, M54,\,\footnote{The case of M54 is notably special. The cluster is still embedded within the disrupting stellar body of the Sagittarius dSph, and therefore its accreted origin is unambiguous. However,  \citet{Bellazzini2008} show that M54 coexists with a distinct nuclear component that can be differentiated in velocity-metallicity phase-space. These authors suggest that they formed independently and  M54 plunged to the central region driven by dynamical friction. Because it is impossible for us to quantify how common this feature is amongst the Virgo nucleated galaxies, here we simply take the view that the two components constitute the NSC of Sagittarius.} and NGC\,2419 in the MW, and Mayall\,II/G1 in M31 are canonical examples for stripped NSC candidates.
Evidence for the nuclear origin picture for these systems includes very high surface mass densities, large internal metallicity spreads and significant elemental abundance variations, unusually high eccentricities, and the presence of kinematic subpopulations \citep{Ibata1994,Ibata1995,Norris1995,Sarajedini1995,Norris1997,Lee1999,Pancino2000,Meylan2001,vandenbergh2004}.
Recently, the lower mass GC M19  has been shown to have an intrinsic abundance spread of $\sigma$[Fe/H] = 0.17 dex, only surpassed by $\omega$Cen and M54 in the Galactic GCS \citep{Yong2016}. 
This is highly suggestive of a nuclear origin as well, and so we include M19 in the list of putative surviving NSCs. 
\new{Finally, \citet{Fuentes-Carrera2008} find that three additional high-velocity dispersion  clusters in M31 exhibit very large abundance spreads, and they are included in our analysis as well (Table\,\ref{tab:galmass}). 
}

We invert our best-fit $M_{NSC}$-\mstar\  relation to infer the stellar masses for the progenitor galaxies of these star clusters under the assumption that they are indeed fully stripped NSCs. 
To do this we first compute the masses of the clusters using $M_{V}$ measurements from the literature, as indicated in the last column of Table\,\ref{tab:galmass}. 
For all clusters we then assume a common $(g-V) = 0.4$ color, and \mstar/$L_{g} = 2.7$ \citep[e.g.,][]{vandeven2006,Noyola2010}.
The second column shows the present-day GC masses, here assumed to be identical to $M_{NSC}$.
The third column corresponds to the expected mean stellar masses for the parent galaxies,  \mstar$_{,p}$, which range from $10^{8}$ \msun\ to  $10^{9}$ \msun.
We recall that the significant intrinsic scatter of the $M_{NSC}$-\mstar\  relation implies that these values can be a factor $\approx$\,2.5 larger or smaller.
In any case, it is clear that only a handful of these (presumably) disrupted systems can contribute $\gtrsim10^{9}$ \msun\ to the stellar halos of the central galaxies in the LG. 

This aligns well with mounting evidence from  both numerical \citep{Bullock2005,Cooper2010,Deason2016,VanOirschot2017} and observational work \citep{Fiorentino2014,Deason2015} that the bulk of the accreted stellar halo in MW-sized galaxies is contributed by a small number of relatively massive satellites with stellar masses $10^{8}-10^{10}$ \msun.
For reference, the MW is thought to have a stellar halo $M_{*,h} \approx 1\times10^{9}$ \msun, corresponding to roughly 2\% of its stellar mass \citep{Carollo2010,Licquia2015}.
M31's halo is slightly more massive, $M_{*,h} \approx 4\times10^{9}$ \msun, amounting to nearly 4\% of its stellar mass  \citep{Courteau2011,Sick2014}.
Comparison with Table\,\ref{tab:galmass} indicates that 30-100\% of the mass in these halos can be accounted for by the  progenitors of (known) NSC candidates. 
\new{This is only a crude comparison and the uncertainties are important--e.g., we have assumed that the galaxy stars are fully stripped, but depending on the orbital configuration a fraction of that material can remain bound and form an 'envelope' around the nucleus \citep[e.g.][]{Pfeffer2013,Liu2015b};}
also, a non-negligible fraction of the material released by a threshed satellite is deposited  in the inner rather than in the outer halo, where stars mix with a pre-existing in situ population \citep{Zolotov2009}.
But if indeed the subset of massive, peculiar GCs in the Local Group are the remnants of disrupted galaxies we propose that  relatively  massive {\it nucleated} satellites constituted a significant fraction of the  building blocks for the stellar halos in the MW and in M31.
A better understanding of the relation between the stellar populations of NSCs and their host galaxies can guide us to identify coherent structures in the multidimensional spatial-kinematical-chemical phase-space originating from material stripped from these galaxies.

\section{Conclusions}
\label{sect:conclusions}

In this study we use deep, high spatial resolution optical imaging from the NGVS to detect and characterize the NSCs in a volume- and mass-limited sample of nearly 400 galaxies in the core of the Virgo cluster spanning seven decades in stellar mass.
Here we have focused on the occurrence of NSCs as a function of galaxy mass and environment, and on mass scaling relations with their host galaxies.
Our main conclusions are as follows: 

\begin{itemize}

\item[1.]{
The NSC occupation distribution is a strong function of galaxy stellar mass. 
It peaks at $f_{n} \approx$ 90\% for \mstar\ $\approx 10^{9}$ \msun\ galaxies, and then declines monotonically for both more and less massive galaxies.
The distribution is shaped by the interplay between the disruptive effects of MBHs at the high mass end, and (possibly) a low initial number of dense star clusters at the low mass end, where $f_{n} \propto \mbox{log}\,M_{*}^{1/4}$.
We identify a characteristic mass \mstar\ $\approx 5\times10^{5}$ \msun\ below which no galaxy in the core of Virgo is nucleated.
}

\item[2.]{
We compare the NSC occupation distribution in Virgo with other environments spanning a wide range of host halo masses, including the Coma and Fornax clusters, and the MW, M31 and M81 groups.
We unveil a secondary dependence of $f_{n}$ on environment, such that at fixed galaxy stellar mass nucleation is more frequent in more massive host halos. 
}

\item[3.]{
NSCs have integrated colors that primarily depend on their stellar mass, such that more massive nuclei  are redder. 
Because $M_{NSC}$ and \mstar\ also are correlated, redder NSCs inhabit redder galaxies--but the poor correlation between these quantities indicates this is only a weaker relation.  
}

\item[4.]{
There is a universal, nonlinear relation between $M_{NSC}$ and \mstar, such that the nucleus-to-galaxy stellar mass ratio drops to $M_{NSC}/$\mstar\ $\approx 3.6 \times 10^{-3}$ for galaxies of mass \mstar\ $\approx 5\times10^{9}$ \msun.
NSCs in both more and less massive galaxies are much more prominent, with the latter  scaling as $M_{NSC} \propto M_{*}^{0.46}$. 
This implies that the faintest nucleated galaxies in the core of Virgo host NSCs that are nearly $50$\% as massive as the galactic body itself. 
However, we also measure an intrinsic scatter in the $M_{NSC}$-\mstar\ relation of 0.4 dex, which we interpret as evidence for stochastic growth of NSCs.
}

\item[5.]{
This universal relation can be inverted to infer the masses for the progenitors of UCDs and massive GCs in the Local Group under the hypothesis that they are the NSCs of tidally disrupted satellites.
From this exercise we conclude that relatively massive nucleated satellites constituted a significant fraction of the building blocks for the stellar halos of $L^{*}$ galaxies.
}

\item[6.]{
We construct the first volume- and mass-limited NSC mass function in Virgo, which peaks at $M_{NSC} \approx 7\times10^{5}$ \msun\ and has a standard deviation of 0.68 dex. 
Comparison with the GCMF indicates that the average NSC in Virgo is 3-4 times more massive than the typical GC.
}

\item[7.]{
We find a close connection between NSCs and GCs, in the sense that the fraction of galaxies hosting either type of star cluster system decreases toward lower masses at the same rate. 
Additionally, the total mass of the GCS is similar to the NSC mass for \mstar\ $\lesssim 10^{9}$ \msun\ galaxies, but the relation breaks down at the high mass end due to a combination of excessively prominent NSCs and an apparent scarcity of GCs. 
}

\item[8.]{
The mass of the NSC exhibits a (logarithmic) linear relation with the estimated peak DM halo mass, but its slope is steeper than the corresponding if the NSC mass fraction were constant, $M_{NSC} \propto M_{h}^{1.2}$.
}

\item[9.]{
Current models for NSC formation including dissipative and dissipationless processes reproduce, at least qualitatively, the observed trends. 
Unfortunately, neither the nucleation fraction nor the $M_{NSC}$-\mstar\ relation seem to have enough discriminative power to distinguish between these scenarios or quantify their relative contribution to NSC formation.
We are however able to show that a model for self-regulated growth of nuclei driven by stellar feedback is not sufficient to explain the observed NSC masses.
We speculate that galactic nuclei formation is best explained by a biased process whereby dense star clusters preferentially form and aggregate in the earliest collapsing halos, and that the  subsequent level of growth is determined by the average nuclear mass infall rate.
}

\end{itemize}

\new{ We find that  \mstar\ $\sim 10^{9.5}$ \msun\ seems to be a very interesting mass scale where (i) the NSC occupation fraction peaks; (ii) the NSC mass fraction reaches a minimum; and (iii) the NSC and GC occupation fractions stop tracking each other. 
This remarkable coincidence is highly suggestive of the existence of an underlying physical mechanism(s) regulating the growth of both NSCs and their host galaxies. 
These three observational results shall inform numerical and theoretical models for the formation of NSCs.
}
In future NGVS papers we will expand the study of NSCs to the entire virial volume of Virgo, where we can investigate whether the occupation fraction depends on clustercentric position or local environmental density. 
We will also extend the studies on NSC occurrence and the relation to their hosts to star forming galaxies, as well as improve the statistics at the high-mass end.
Constructing a large sample of NSCs in galaxies in the $10^{9}-10^{11}$ \msun\ range is the next critical step to fully understand the nature of the rare and extended NSCs that produce the bend in the $M_{NSC}$-\mstar\ relation. 
Finally, we plan to exploit the multiwavelength $u'griz'$ photometry provided by the NGVS to carry out studies of their stellar population content through modelling of their SEDs \citep[e.g.,][]{Spengler2017}.


\acknowledgments
The authors acknowledges Fabio Antonini for providing the data from his semianalytic models in electronic format. 
RSJ would like to thank Diederik Kruijssen, Nadine Neumayer, Joel Pfeffer and Anil Seth  for useful discussions and suggestions.
The authors acknowledge an anonymous referee for a thorough and constructive review of the manuscript. 
THP acknowledges support by the FONDECYT Regular Project Grant (No. 1161817) and the BASAL Center for Astrophysics and Associated Technologies (PFB-06). 
S.M. acknowledges financial support from the Institut Universitaire de France (IUF), of which she is senior member. 
This work is supported in part by the Canadian Advanced Network for Astronomical Research (CANFAR) which has been made possible by funding from CANARIE under the Network-Enabled Platforms program. This research used the facilities of the Canadian Astronomy Data Centre operated by the National Research Council of Canada with the support of the Canadian Space Agency. The authors further acknowledge use of the NASA/IPAC Extragalactic Database (NED) which is operated by the Jet Propulsion Laboratory, California Institute of Technology, under contract with the National Aeronautics and Space Administration.

\bibliography{library.bib}

\begin{appendices}
\section{NSCs in the MW, M31 and M81 systems}
\label{sect:appendix}

In \myfig{fig:frac} we include two datapoints  corresponding to the nucleation fraction in the  MW, M31, and M81 groups.
As has been broadly discussed in the literature, identifying robustly identifying NSCs is in many cases far from trivial. 
Here we detail the process we have followed in this work to perform the nucleation classification for nearby satellites.
The interested reader is referred to Sect.\,4.2 in \citet{Turner2012} for a more detailed discussion on other systems that feature structural and kinematical peculiarities in their nuclear regions, but that we do not classify as nucleated.

We select all candidates from the \citet{Karachentsev2013} Updated Nearby Galaxy Catalog.
\new{We transform the listed $B$-band magnitudes to the $V$-band assuming $(B-V)=0.7$ and compute stellar masses using \mstar/$L_{V}=1.6$ \citep{Woo2008}.}
We further select only galaxies with stellar masses in the $10^{5} < M_{*}/M_{\odot} < 10^{9}$ range with early-type  morphologies ($T < 0$) and whose main perturber is either of the three central spirals.
The final sample consists of 55 nearby satellites (Table \ref{tab:lv_nuc}).

Of these, only one satellite galaxy is considered to be nucleated in the MW system, namely the Sagittarius dSph \citep{Mateo1998,Monaco2005}. 
In M31 we count M32 and NGC\,205 as unambiguously having NSCs.
Finally, for satellites in the M81 group we have visually inspected the {\it HST} images for all the candidate galaxies from \citet{Chiboucas2013}.
Only two systems, KDG\,61 and KDG\,64 are identified as nucleated.




\end{document}